\documentclass[12pt]{article}

\pdfoutput=1

\usepackage{ifpdf}
\ifpdf
\usepackage{graphicx,color}
\usepackage{hyperref}
\else
\usepackage[dvipdfmx]{graphicx,color}
\usepackage[dvipdfmx]{hyperref}
\fi
\usepackage{amssymb, amsfonts, amsmath, cancel, cite, multirow}
\numberwithin{equation}{section}
\usepackage{mathrsfs}
\usepackage{tikz}
\usepackage[capitalise]{cleveref}
\usepackage[margin=1in]{geometry}
\usepackage{braket}
\usepackage{ulem}
\begin{document}

\begin{titlepage}

\begin{flushright}
\end{flushright}

\vskip 1.35cm
\begin{center}

{\large
\textbf{
   Unitarization of the Sommerfeld enhancement through the renormalization group
}
}
\vskip 1.2cm

\renewcommand{\thefootnote}{\fnsymbol{footnote}}
Yuki Watanabe\footnote{Email: yuki.watanabe@ipmu.jp}
\renewcommand{\thefootnote}{\arabic{footnote}}

\vskip 0.4cm

\textit{
    Kavli IPMU (WPI), UTIAS,
    University of Tokyo, Kashiwa, Chiba 277-8583, Japan
}

\vskip 1.5cm
\begin{abstract}
    When a pair of dark matter particles interacts via a long-range force mediated by a light particle, their nonrelativistic annihilation cross section can be significantly enhanced—a phenomenon known as the Sommerfeld enhancement. This enhancement exhibits resonant behavior if the long-range potential supports shallow bound states or narrow resonances, which can lead to violations of the partial-wave unitarity bound. We identify the origin of this pathological behavior as the emergence of secular terms in perturbative expansions associated with low-energy composite states of the long-range potential. To address this issue, we propose a renormalization group improvement of the perturbative series. The resulting improved amplitude provides a unitarity-consistent form of the Sommerfeld enhancement, with its poles acquiring an imaginary part that reflects the decay width of the annihilating bound states. We also briefly discuss the implications of our approach from the perspective of Wilsonian renormalization group, and comment on its potential application to higher-order annihilation processes such as bound-state formation.

\end{abstract}
\end{center}
\end{titlepage}
\section{Introduction}

Dark matter (DM) constitutes a major component of our universe, and its nature is deeply intertwined with the history of cosmic evolution. However, our current understanding of DM — namely, the quantum-mechanical aspects of its nature — remains extremely limited; quantitatively, we only have information regarding its mass range and relic abundance in our universe\,\cite{Cirelli:2024ssz}. Given that the Standard Model of particle physics remains incomplete, DM must be incorporated as a crucial missing piece. While its evidence lies in cosmology and astrophysics, DM thus may serve as a portal to the innermost structure of particle physics. Unraveling the enigma of dark matter is essential to closing the Ouroboros loop that connects the microscopic world of particle physics with the macroscopic scale of the cosmos.

When DM plays an essential role in the time evolution of the universe, it is non-relativistic and cold. This constitutes a key feature of DM, which in turn has nontrivial implications for the construction of viable DM models. In the presence of a light mediator particle inducing long-range interactions between dark matter particles, the annihilation cross section can be significantly enhanced at low velocities—a phenomenon known as the Sommerfeld effect or Sommerfeld enhancement\,\cite{Sommerfeld:1931qaf,Hisano:2002fk, Hisano:2003ec, Hisano:2004ds, Arkani-Hamed:2008hhe}. Taking Sommerfeld enhancement into account is crucial for model evaluations, especially in calculating the relic abundance and deriving constraints from indirect detection.

Sommerfeld enhancement exhibits a resonant behavior and becomes significantly amplified when the parameters of the long-range force approach certain critical values. This occurs when the spectrum generated by the long-range interaction contains shallow bound states or narrow resonances\,\cite{Hisano:2004ds, Kamada:2023iol, Beneke:2024iev}. However, such an excessive amplification of the annihilation cross section is, in fact, inconsistent with quantum theory. Specifically, it violates the unitarity of quantum mechanics—more precisely, it leads to a contradiction with the unitarity bound on the annihilation cross section, which is a direct consequence of unitarity\,\cite{Hisano:2004ds, Kamada:2023iol, Blum:2016nrz}. For example, in the case of $s$-wave annihilation, if there exists a shallow bound state, the annihilation cross section including the Sommerfeld enhancement scales as $v^{-2}$ with respect to the dark matter velocity $v$, whereas the unitarity bound scales as $v^{-1}$, leading to an inconsistency in the small-$v$ regime. Accepting the excessively enhanced Sommerfeld effect without modification may result in a misrepresentation of the model’s validity. Accordingly, it is of crucial importance to unitarize the Sommerfeld enhancement—namely, to reformulate it in a manner consistent with unitarity.

The unitarization of the Sommerfeld enhancement has been addressed in several previous works\,\cite{Blum:2016nrz, Parikh:2024mwa, Flores:2024sfy}. The essential idea underlying these approaches, as already pointed out in the original work\,\cite{Hisano:2004ds}, is to properly resum the short-range effects that encode the annihilation processes. For instance, Ref.\,\cite{Blum:2016nrz} implements this resummation by formulating the short-range contributions as a potential and solving the Schr\"{o}dinger equation accordingly. However, the short-range potentials introduced in such formulations are highly singular, requiring careful treatments\,\cite{Lepage:1997cs}.

Incorporating short-distance physics via effective potentials is a well-established and legitimate strategy, often employed in nuclear physics when analyzing nuclear forces\,\cite{Kaplan:1996nv, Kaplan:1996xu, Kaplan:1998we}. Yet, this method does not directly explain the origin of the pathological enhancement seen in the Sommerfeld enhancement. In the case of nuclear interactions, strong coupling naturally invalidates the Born approximation. But for dark matter annihilation—which is typically weakly coupled at the UV scale—the failure of perturbative calculations calls for a deeper explanation. While one might argue that the Sommerfeld enhancement effectively renders the annihilation process strongly coupled, such reasoning alone does not fully account for the emergence of nonperturbative behavior in otherwise perturbative settings. Thus, although the two systems may share formal similarities, the underlying structures and mechanisms responsible for nonperturbativity are fundamentally distinct. Furthermore, Ref.~\cite{Hisano:2004ds} also points out that the actual Sommerfeld enhancement is regulated by the decay width associated with shallow bound states, thereby ensuring consistency with unitarity. However, existing approaches based on solving the Schr\"{o}dinger equation do not provide a clear or direct account of how this regulation mechanism emerges.

In this work, we investigate the violation of unitarity in the Sommerfeld enhancement within a minimal setting in which DM is described by a single-state particle that interacts via a long-range force and undergoes annihilation through a perturbative short-range interaction. We identify the origin of the unitarity violation as the emergence of secular terms in the perturbative expansion, arising in the presence of shallow bound states or narrow resonances. That is, the standard perturbative treatment used to derive the Sommerfeld enhancement lacks uniform convergence in momentum space, owing to an infrared divergence arising from the spectral structure induced by the long-range potential; in this sense, it represents a case of singular perturbation theory.

Singular perturbation theory is a well-known issue, and it is established that it can be systematically addressed using renormalization group methods\,\cite{Chen:1994zza, Chen:1995ena}. In this paper, we propose a unitarization scheme for the Sommerfeld enhancement, in which we resum the secular terms that appear in the perturbative expansion of the self-scattering amplitude using the renormalization group approach. The ingredients required for this procedure are the short-distance scattering amplitude from the UV theory and the phase shift induced by the long-range interaction. The resulting improved scattering amplitude, as well as the modified Sommerfeld factor derived from it, are both manifestly consistent with unitarity. Furthermore, by analyzing its analytic structure, we confirm the appearance of an imaginary part in the bound-state energy—i.e., a decay width—thus realizing the physical interpretation proposed in Ref.~\cite{Hisano:2004ds} in a concrete manner.

The organization of this paper is as follows. In the next section (Section~\ref{sec:the standard SE}), we first introduce the basic tools of nonrelativistic scattering theory, which are then used to derive the conventional Sommerfeld enhancement. Through this derivation, the origin of the unitarity violation becomes manifest. In the following Section~\ref{sec: RG improve}, motivated by this breakdown of unitarity, we introduce a renormalization group approach. We then analyze the properties of the resulting improved scattering amplitude, confirming that it is consistent with unitarity and that the positions of its poles are shifted to include the decay widths of the corresponding bound states.
Subsequently, in Section~\ref{sec: Discussion}, we employ insights from Wilsonian renormalization group theory to further discuss our results. We also comment on the approach based on directly solving the Schr\"{o}dinger equation found in\,\cite{Blum:2016nrz,Parikh:2024mwa}. A brief discussion is also given on bound-state formation\,\cite{An:2016gad,Petraki:2015hla,Petraki:2016cnz} as a higher-order annihilation process.
Section~\ref{sec: Conclusion} is devoted to the conclusion. Appendix ~\ref{app: Bessel} provides a summary of the Bessel functions used in the nonrelativistic scattering theory, and Appendix~\ref{app: Jost} collects technical details omitted in the main text. Appendix~\ref{app: numerical} numerically confirms that our method reproduces the results of \cite{Blum:2016nrz,Parikh:2024mwa}.
\section{The conventional Sommerfeld enhancement}
\label{sec:the standard SE}

We discuss the conventional Sommerfeld enhancement within the framework of the scattering theory in non-relativistic quantum mechanics\,\cite{Newton:1982qc,Taylor:1972pty,Weinberg:1995mt}. After introducing the foundational elements of the scattering theory, we proceed to the analytic structure of the scattering amplitude in non-relativistic systems, placing particular emphasis on the role of the Jost function\,\cite{Taylor:1972pty,Hyodo:2020czb}, which also plays a central role in the formulation and understanding of the Sommerfeld effect. Although the development presented here may initially appear somewhat unconventional and exaggerated, one will find that it is an essential component in the unitarization of the Sommerfeld effect, which is the central subject of this paper. Although our main focus is on scattering phenomena involving DM, we will refrain from specifying the identity of the scattering particles throughout the rest of this paper.

\subsection{Basic concepts in scattering theory}

We begin with the most fundamental quantity in scattering theory: the $S$-matrix. Consider a pair of distinguishable quantum mechanical particles with reduced mass $\mu$.\footnote{In the case of identical particles, additional numerical factors (such as 2) may appear in certain expressions\,\cite{Flores:2024sfy,Parikh:2024mwa}.} These particles evolve under the free Hamiltonian $H_0$ in the distant past, interact via an interaction $V$ localized around time $t = 0$, and then continue to evolve freely under $H_0$ in the distant future. The $S$-matrix element for the transition from an asymptotically free state to another, labeled by $\alpha$ and $\beta$, is defined as
\begin{align}
    S_{\beta \alpha} = \braket{\beta^{-}|\alpha^{+}}.
    \label{eq:def of the s-matrix}
\end{align}
Here, the in (or out) asymptotic state $\ket{\alpha^{\pm}}$ represents the physical state at time $t = 0$ that asymptotically approaches the eigenstate $\ket{\alpha}$ of the free Hamiltonian $H_0$—with energy $E_\alpha$ and momentum $\vec{p}_\alpha$—in the infinite past (or future) and satisfies the Lippmann–Schwinger equation:
\begin{align}
    \ket{\alpha^{\pm}} = \ket{\alpha} + \frac{1}{E_\alpha - H_0 \pm i0^+} V \ket{\alpha^{\pm}}.
\label{eq:LS equation}
\end{align}
The unitarity of the $S$-matrix, $\sum_\gamma S_{\alpha \gamma} S^\dagger_{\gamma \beta} = \sum_{\gamma}S^\dagger_{\alpha \gamma} S_{\gamma \beta} = \delta_{\alpha \beta}$, follows from its definition in Eq.~(\ref{eq:def of the s-matrix}) and the Lippmann-Schwinger equation (\ref{eq:LS equation}). It should be noted that the asymptotic states $\ket{\alpha^\pm}$ are eigenstates of the total Hamiltonian $H=H_0+V$ with eigenvalue $E_\alpha$.

It is convenient to define the scattering amplitude denoted by $f(\vec{p}_\alpha \to \vec{p}_\beta)$, which isolates the nontrivial part of the $S$-matrix by subtracting the identity contribution. This leads to the expressions
\begin{align}
    S_{\beta \alpha} = \delta^{(3)}(\vec{p}_\beta-\vec{p}_\alpha) + \frac{i}{2\pi \mu} \delta(E_\beta - E_\alpha) f(\vec{p}_\alpha \to \vec{p}_\beta), \quad \frac{d\sigma}{d\Omega} = |f(\vec{p}_\alpha \to \vec{p}_\beta)|^2,
\end{align}
where the first equation shows how the amplitude $f$ enters the $S$-matrix, and the second the differential cross section, indicating its physical significance. The Lippmann-Schwinger equation (\ref{eq:LS equation}) provides a simple and exact form of the scattering amplitude
\begin{align}
    f(\vec{p}_\alpha \to \vec{p}_\beta)
    =-(2\pi)^2\mu \braket{\beta|V|\alpha^+}.
\end{align}

When the Hamiltonian possesses rotational symmetry, it is natural to expand the asymptotic states in the angular momentum basis that simultaneously diagonalizes the Hamiltonian $H$, the total angular momentum $J^2$, and its third component $J^3$. We denote these states $\ket{E\ell m^\pm}$, where $E$, $\ell$ and $m$ are the eigenvalues of $H$, $J^2$, and $J^3$ respectively. On this basis, the $S$-matrix becomes diagonal, and its matrix elements take the form
\begin{align}
    \braket{E' \ell' m'^- | E \ell m^+} = \delta(E - E') \delta_{\ell' \ell} \delta_{m' m} e^{2i\delta_\ell(p)},
\end{align}
where $\delta_\ell(p)$ is the phase shift for the angular momentum $\ell$ and is a real function due to the unitarity of the $S$-matrix. Here, we parametrize the energy $E$ using the magnitude of the momentum $E=p^2/2\mu$. The partial-wave decomposition of the scattering amplitude is then given by
\begin{align}
    \label{eq:PWA}
    f(p, \theta) = \sum_{\ell = 0}^\infty (2\ell + 1) f_\ell(p) P_\ell(\cos\theta), \quad f_\ell(p) = \frac{e^{2i\delta_\ell(p)} - 1}{2ip} = \frac{p^{2\ell}}{p^{2\ell + 1} \cot \delta_\ell(p) - ip^{2\ell + 1}},
\end{align}
where $P_\ell(x)$ is the Legendre polynomial of degree $\ell$, $\theta$ is the scattering angle, and $f_\ell(p)$ is the $\ell$-th partial wave amplitude. The total cross section $\sigma(p)$ can also be expressed as a sum over the partial-wave contributions:
\begin{align}
    \sigma(p) = \sum_{\ell = 0}^\infty \sigma_\ell(p), \quad
    \sigma_\ell(p) = 4\pi (2\ell + 1)\, |f_\ell(p)|^2 = \frac{4\pi (2\ell + 1)}{p^2} \sin^2 \delta_\ell(p).
\end{align}

The unitarity of the $S$-matrix imposes nontrivial constraints on the form of the scattering amplitude, which are referred to as the optical theorem. In the case where multiple scattering channels are present, the condition of unitarity on the partial-wave amplitudes $f_{\ell \,ij}$, where $i$ and $j$ label the incoming and outgoing channels respectively, leads to the following relation\,\cite{Oller:2020guq, ParticleDataGroup:2024cfk}:
\begin{align}
    \label{eq:optical theorem}
    \operatorname{Im} f_{ij} = \sum_k p_k f_{ik} f_{jk}^*,
\end{align}
where the summation is taken over all open channels $k$, with $p_k$ denoting the corresponding channel momentum. For a single channel scattering (namely, elastic scattering), this reduces to a simpler form,
$\operatorname{Im} f(p) = p |f(p)|^2$,
which is consistent with Eq.~(\ref{eq:PWA}).

An important concept in scattering theory is the analyticity of the scattering amplitude, which is considered to follow from causality\,\cite{Mizera:2023tfe,Hannesdottir:2022bmo}. That is, the scattering amplitude can be analytically continued in terms of appropriate kinematic variables lifted to the complex plane, and the original physical scattering information is recovered as the boundary value of these variables. For example, a partial-wave scattering amplitude $f_\ell(E)$ is analytically continued with energy 
$E$ regarded as a complex variable, and the physical amplitude in $E>0$ is given by $f_\ell(E+i0)$.

One of the major utilities of the analyticity of the amplitude lies in the statement that its singularities are believed to correspond to spectral properties of the system. Specifically, bound states correspond to poles of the scattering amplitude. Moreover, unitarity, or equivalently the optical theorem, suggests that the scattering amplitude possesses branch cuts with thresholds as branch points. Indeed, requiring Hermitian analyticity of the amplitude, $f_\ell(E^*)=[f_\ell(E)]^*$ leads, via the optical theorem, to the presence of an imaginary part in the scattering region, which in turn results in discontinuities and hence the presence of branch cuts. In the next section, we will see explicitly how, in non-relativistic quantum mechanics, the $S$-matrix or scattering amplitude can be analytically continued into complex energy or complex momentum, and realize these structures.

\subsection{The Jost function and the analytic properties of the $S$-matrix}

We focus on the single-channel scattering problem. In the non-relativistic limit, the Lippmann-Schwinger equation (\ref{eq:LS equation}) can be expanded in terms of position eigenstates $\ket{\vec{r}}$ and reduced to an integral equation, which then can be transformed into an equivalent differential equation, the Schr\"{o}dinger equation for the scattering state $\braket{\vec{r}|\vec{p}^{+}} = \psi^{+}_p(\vec{r})$.\footnote{Due to time-reversal symmetry, the scattering wave function for the out-state $\psi^-_p(r)$ is related to that of the in-state $\psi^+_p(r)$ by complex conjugation.} Assuming a spherically symmetric potential $V(\vec{r})=V(r)$, the wave function $\psi^{+}_p(\vec{r})$ can be expanded by the spherical waves as
\begin{align}
    \psi_p^+(\vec{r}) = (2\pi)^{-3/2}\frac{1}{pr}\sum_{\ell=0}^\infty (2\ell+1) i^\ell  u_{\ell,p} (r)P_\ell(\hat{\vec{r}}\cdot\hat{\vec{p}}),
\end{align}
where the radial wave function $u_{\ell,p}(r)$\footnote{The radial wave function $u_{\ell,p}(r)$ is normalized as $\int^\infty_0 dr\,u^*_{\ell,p}(r) u_{\ell,p'}(r) = \pi \delta(p-p')/2$.} follows the radial Schr\"{o}dinger equation with its eigen-energy being given by $E = p^2/2\mu$,
\begin{align}
    \label{eq:radial sceq}
    \left[
        -\frac{1}{2\mu}\frac{d^2}{dr^2} 
        + V(r)
        + \frac{\ell(\ell+1)}{2\mu r^2}
    \right]
    u_{\ell,p}(r)
    =
    E u_{\ell,p}(r).
\end{align}
The solution of (\ref{eq:radial sceq}) we need for the scattering problem is completely determined by its boundary conditions and they are, at the origin, the regularity of the wave function $u_{\ell,p}(0) = 0$ \footnote{The irregular solution implies that the Hamiltonian includes a delta-function potential localized at the origin\,\cite{Newton:1982qc}. Imposing the regularity condition excludes the existence of such a singular operator. Conversely, for a Schr\"{o}dinger equation with a singular potential, the solution necessarily contains an irregular component.}and, at infinity,  that the wave function is a superposition of an incoming wave and an outgoing wave,
\begin{align}
    \label{eq:scat bc}
    u_{\ell,p}(r) \to \frac{i}{2}[h_\ell^{-}(pr)-s_\ell(p)h^{+}_\ell(pr)] \quad (r \to \infty),
\end{align}
where $h_\ell^{\mp}(pr)$ are the Riccati-Hankel functions, which are the solutions of the Schr\"{o}dinger equation (\ref{eq:radial sceq}) with $V(r)=0$. Their asymptotic behavior at infinity are given by $h_\ell^\mp(pr) \sim e^{\mp ipr}$, that is, $h^-_\ell(pr)$ and $h^+_\ell(pr)$ correspond to the incoming and outgoing waves, respectively (See Appendix~\ref{app: Bessel}). Therefore, in the formulation of the scattering problem based on the Schr\"{o}dinger equation, the $S$-matrix is extracted from the relative weight of the incoming and outgoing waves. 

We investigate the analytic continuation of the $S$-matrix to the complex $p$-plane.
The scattering problem under a central potential $V(r)$ reduces to solving the Schr\"{o}dinger equation (\ref{eq:radial sceq}) as a Dirichlet problem with boundary conditions at the origin and at infinity.
However, when one aims to explore the  scattering theory at a deeper level, including the general spectral properties of a quantum system, such as bound states and resonances, it is often highly convenient to study the Schr\"{o}dinger equation under boundary conditions different from those we discussed above. Here, we introduce the regular solution of Eq.~(\ref{eq:radial sceq}), $\phi_{\ell,p} (r)$, defined as the solution to the Cauchy problem specified by the following boundary condition:
\begin{align}
    \label{eq:regular bc}
    \phi_{\ell,p}(r)/j_\ell(pr) \to 1 \quad (r \to 0).
\end{align}
This condition fixes both the value of the function and its derivative at the origin. The normalization of the regular solution $\phi_{\ell, p}(r)$ is thus determined by the boundary condition in (\ref{eq:regular bc}), and is proportional to the Riccati-Bessel function $j_\ell(pr)$, which satisfies Eq.~(\ref{eq:radial sceq}) with vanishing potential $V(r) = 0$, and behaves as $j_\ell(pr) \to (pr)^{\ell+1}/(2\ell+1)!!$ near the origin (See Appendix~\ref{app: Bessel}).

We now examine the properties of the regular solution $\phi_{\ell,p}(r)$, focusing on its role in the construction of the analytic structure of the $S$-matrix $s_\ell(p)$ or the scattering amplitude $f_\ell(p)$. Note that in the Dirichlet problem, the equation admits a solution under given boundary conditions only for certain values of $p$, such as $p>0$ for the scattering boundary condition (\ref{eq:scat bc}), while as a Cauchy problem, the equation admits a solution for complexified momentum $p$. This fact naturally gives the analytic continuation of the $S$-matrix $s_\ell(p)$ into the complex $p$ plane.\footnote{The region in the complex $p$-plane to which the $S$-matrix can be analytically continued depends on the form of the potential.} If the potential $V(r)$ decays sufficiently rapidly at infinity and the centrifugal potential becomes dominant, then the regular solution can be expressed, at large distances, as a linear combination of Riccati–Hankel functions $h_\ell^{\mp}(pr)$; that is,
\begin{align}
    \label{eq:reg asymptotic}
    \phi_{\ell,p}(r) \to \frac{i}{2}[\mathscr{J}_\ell(p)h^-_\ell(pr)-\mathscr{J}_\ell(-p)h^+_\ell(pr)].
\end{align}
Here, the coefficient $\mathscr{J}_\ell(p)$ of the incoming wave is called the Jost function, and it admits an integral representation in terms of the regular solution as\footnote{Formally, the result is first established for real positive $p>0$, and then analytically continued to complex $p$. For the derivation of this result, see Appendix~\ref{app: Jost}.}
\begin{align}
    \label{eq:def jost}
    \mathscr{J}_\ell(p) = 1 + \frac{2 \mu}{p} \int^\infty_0 dr h^{\pm}_\ell(pr) V(r) \phi_{\ell,p}(r).
\end{align}
The integral part of the above equation is bounded from above and one can show that $\mathscr{J}_\ell(\infty) =1$.

We are now in a position to analytically continue the 
$S$-matrix, originally defined for the physical region $p>0$, into the complex 
$p$-plane. Following the boundary condition (\ref{eq:scat bc}) imposed on the scattering wave function, we define the 
$S$-matrix $s_\ell(p)$ for complex 
$p$ as the relative coefficient between the Riccati–Hankel functions and the scattering amplitude $f_\ell(p) = (s_\ell(p)-1)/2ip$:
\begin{align}
    \label{eq:analytic s matrix}
    s_\ell(p) = \frac{\mathscr{J}_\ell(-p)}{\mathscr{J}_\ell(p)}, \quad f_\ell(p) = \frac{\mathscr{J}_\ell(-p)-\mathscr{J}_\ell(p)}{2ip \mathscr{J}_\ell(p)}.
\end{align}
Since both the regular solution $\phi_{\ell,p}(r)$ and the scattering wave function $u_{\ell,p}(r)$ are regular at the origin, they must be proportional to each other in the physical region $p>0$, where the scattering solution $u_{\ell,p}(r)$ exists, and we can find that those are related by $u_{\ell,p}(r)=\phi_{\ell,p}(r)/\mathscr{J}_\ell(p)$ by comparing their asymptotic forms (\ref{eq:scat bc}) and (\ref{eq:reg asymptotic}).
Therefore, the above definition agrees with the original $S$-matrix on the positive real axis and provides its natural analytic continuation into the complex
$p$ plane. In addition, using the relation $\mathscr{J}_\ell(p) = [\mathscr{J}_\ell(-p^*)]^*$, which follows from (\ref{eq:def jost}) (see Appendix~\ref{app: Jost} for details), we obtain $s_\ell(p) =e^{2i\delta_\ell(p)}= \mathscr{J}_\ell^*(p)/\mathscr{J}_\ell(p)$, indicating that the phase of the Jost function in the physical region $p>0$ is given by $-\delta_\ell(p)$.\footnote{There is an ambiguity of $2\pi i$ in the phase shift $\delta_\ell(p) $, and here we choose it to be $\delta_\ell(\infty) = 0$. However, this choice does not affect the subsequent discussion.}

It is noteworthy to consider the case where the analytically continued $S$-matrix or scattering amplitude (\ref{eq:analytic s matrix}) has poles in the complex $p$-plane. Suppose that the $S$-matrix $s_\ell(p) = \mathscr{J}_\ell^*(p)/\mathscr{J}_\ell(p)$ has a pole on the positive imaginary axis of the $p$-plane; that is, the Jost function vanishes at some $p=i\kappa, \kappa>0$.
In such a situation, as can be seen in equation (\ref{eq:reg asymptotic}), the incoming wave component of the regular solution $h_\ell^-(pr)$ vanishes at infinity, while the outgoing wave component remains in decaying form: $\phi_{\ell,p}(r) \sim e^{-\kappa r}$.
This is precisely the boundary condition imposed on the Schr\"{o}dinger equation (\ref{eq:radial sceq}) when solving for bound states.
Conversely, for any value of $p$ where a bound-state wave function exists, it must be proportional to the regular solution due to regularity at the origin, and thus the growing incoming wave component at infinity must be absent.
Therefore, a zero of the Jost function on the positive imaginary axis, that is, a pole of the $S$-matrix or the scattering amplitude\footnote{The zeros of the Jost function do not correspond one-to-one with the poles of the $S$-matrix. This is because some poles may arise from divergences in the numerator, and such poles are referred to as redundant poles. In fact, the $S$-matrix for an exponential potential includes redundant poles.}, corresponds to a bound state, and the associated energy $E=-\kappa^2/2\mu$ gives the binding energy of that state.

On the other hand, the $S$-matrix may have a pole at $ p = p_R - i\gamma $, with $ p_R, \gamma > 0$, corresponding to a zero of the Jost function at that complex value of
$p$. This condition is equivalent to the Gamow–Siegert boundary condition for the Schr\"{o}dinger equation (\ref{eq:radial sceq}), which requires the wave function to behave as a purely outgoing wave at spatial infinity, and although it does not lead to a square-normalizable wave function, it exhibits a physically significant effect. That is, such a boundary condition defines a resonance state, whose associated complex energy is given by
\begin{align}
    E = \frac{p^2}{2\mu} = \frac{p_R^2 - \gamma^2}{2\mu} - i \frac{p_R \gamma}{\mu},
\end{align}
where the real part represents the resonance energy and the imaginary part determines the decay width, $\Gamma = 2p_R\gamma/\mu$.
This complex energy leads to the time dependence of the wave function $\psi(t) \sim e^{-iE_R t} e^{-\Gamma t/2}$, indicating an exponential decay of the amplitude over time.
In fact, the rigorous meaning of such complex energy states is captured by expanding the Jost function in Eq.~(\ref{eq:analytic s matrix}) near the resonance zero in (\ref{eq:analytic s matrix}), which yields the Breit-Wigner formula, a standard representation of the resonant behavior of the cross section.

\begin{figure}[t]
    \centering
    \includegraphics[width=0.95\linewidth]{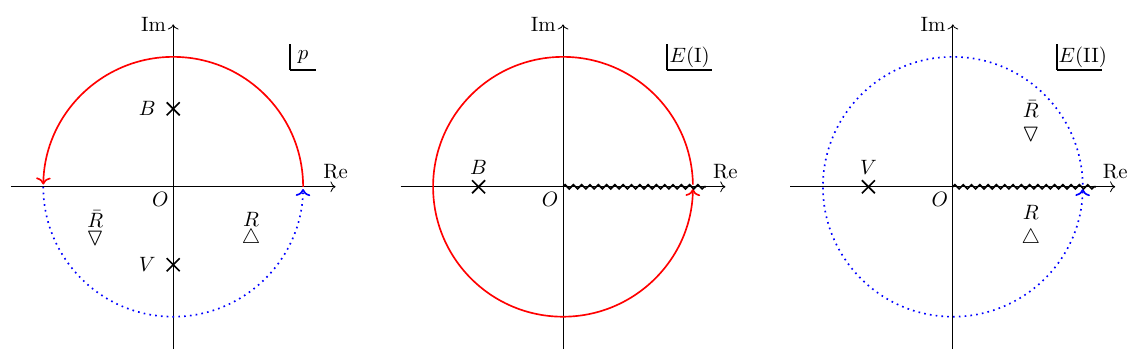}
    \caption{\small \sl 
    Analytic structure of the scattering amplitude on the complex $p$ and $E = p^2/2\mu$ plane.
    The left panel shows the complex $p$-plane, while the middle and right panels represent the physical and unphysical sheets of the energy Riemann surface, respectively. The two sheets are connected across the branch cut, depicted by the wavy lines along the positive real axis. Sample paths illustrating analytic continuation around the branch point, as well as the corresponding trajectories in the $p$-plane and possible pole singularities, are also indicated. See the main text for details.
    }
    \label{fig: Riemann}
\end{figure}

The analytic structure of the scattering amplitude and the Jost function is naturally described on both the complex $p$-plane and the associated Riemann surface of the energy $E=p^2/2\mu$, a representative example of which is illustrated in Fig.~\ref{fig: Riemann}, where the upper (lower) half of the complex $p$-plane corresponds to the first (second) Riemann sheet of the energy, denoted $E({\rm I})$ and $E({\rm II})$. In order to analytically connect these sheets, a branch cut —usually referred to as the right-hand cut or the unitarity cut — is placed along the positive real axis of the complex $E$-plane, consistent with the requirement that scattering occurs for $p>0$ and with the constraint imposed by the optical theorem.\footnote{Depending on the potential, the scattering amplitude may exhibit a left-hand cut, known as a dynamical cut, in contrast to the right-hand cut required by unitarity. For example, the Yukawa potential $V(r) = \alpha e^{-m_Vr}/r$ generates such a cut starting from $E_L =-m_V^2/8\mu$. In relativistic scattering, due to crossing symmetry, the unitarity cut automatically implies the existence of the dynamical cut.} A sample path encircling the branch point at $E=0$ is also shown in Fig.~\ref{fig: Riemann}. The path starts on the upper edge of the branch cut on $E({\rm I})$, moves to the lower edge via the red arc, continues to the upper edge of $E({\rm II})$, and then to its lower edge along the blue dotted arc, finally returning to the starting point on $E({\rm I})$. The corresponding loop is also illustrated on the complex $p$-plane. The physical scattering amplitude is then identified with the boundary value on the upper (lower) edge of the branch cut on $E({\rm I})$\,($E({\rm II})$), and the two sheets are thus referred to as the physical and unphysical sheets, respectively.

As previously discussed, the poles of the scattering amplitude correspond to the system's spectrum. A pole on the positive imaginary axis of the $p$-plane or equivalently, on the negative real axis of the physical sheet $E({\rm I})$, represents a bound state.\footnote{Due to probability conservation, the scattering amplitude cannot have any poles on the physical sheet other than those corresponding to bound states.} On the other hand, resonance poles can appear below the positive real axis of the unphysical sheet $E({\rm II})$. Note that, due to the symmetry properties of the Jost function, $\mathscr{J}_\ell(p) = [\mathscr{J}_\ell(-p^*)]^*$, a resonance pole $p=p_{\rm res}$ is always accompanied by its "conjugate" pole at $p = -p^*_{\text{res}}$, usually referred to as the antiresonance. However, unlike resonance poles, antiresonance poles are located far from the unitarity cut and thus have little effect on the scattering process. The amplitude also admits virtual states—poles on the negative imaginary axis of the $p$-plane that reside on the unphysical sheet $E({\rm II})$. These poles of the scattering amplitude —a bound state, a resonance, an antiresonance, and a virtual state— are indicated in Fig.~\ref{fig: Riemann} as $B$, $R$, $\bar{R}$ and $V$, respectively.

Analyzing the behavior of the Jost function around $p = 0$ is crucial for the investigation of low-energy scattering. From Eq.~(\ref{eq:def jost}), it can be shown that, the Jost function is analytic at $p=0$ and admits the following expansion:
\begin{align}
    \label{eq:low energy expansion of jost}
    \mathscr{J}_\ell(p) = F_\ell(p^2) + ip^{2\ell+1}G_\ell(p^2),
\end{align}
where $F_\ell(p^2)$ and $G_\ell(p^2)$ are some real analytic functions (See Appendix~\ref{app: Jost} for its derivation). Note that, in terms of energy $E=p^2/2\mu$, both functions $F_\ell$ and $G_\ell$ are also analytic in $E$; The branch point singularity of $\mathscr{J}_\ell$ as a function of $E$ arises from the square root $p = \sqrt{2\mu E}$ multiplying $G_\ell$. By combining the expansion of the Jost function (\ref{eq:low energy expansion of jost}) and the formula for the scattering amplitude (\ref{eq:analytic s matrix}), one can derive the effective range expansion that characterizes the energy dependence of the non-trivial part of the scattering amplitude at low momenta,
\begin{align}
    \label{eq:effective range expansion}
    p^{2\ell+1}\cot \delta_\ell(p) = -\frac{1}{a_\ell} + \frac{r_\ell}{2}p^2
    + \mathcal{O}(p^4),
\end{align}
where the parameters $a_\ell$ and $r_\ell$ are the so-called the scattering length and the effective range of the $\ell$-th partial wave, respectively. In particular, the approximation of the scattering amplitude by two parameters, $a_\ell$ and $r_\ell$, is referred to as the effective range approximation\,\cite{Bethe:1949yr}. The effective range expansion obtained above is well motivated by the analyticity of the scattering amplitude as a function of the energy $E$. While the optical theorem, ${\rm Im} \, f_\ell^{-1}(p) = -p$, determines the singular part associated with the unitarity cut of the inverse amplitude, apart from the kinematic singularity associated with the low-energy behavior $f_\ell(p) \propto p^{2\ell}$, the real part ${\rm Re} \, p^{2\ell}f^{-1}_\ell(p) = p^{2\ell+1}\cot \delta_\ell(p)$ is analytic in
$E$, justifying its expansion in powers of $p^2$ near the threshold.

\subsection{Sommerfeld enhancement}
\label{sec:Sommerfeld enhancement}
We now consider a scattering problem that includes both elastic and inelastic processes, and discuss the Sommerfeld effect in the annihilation process using the concepts introduced so far. The description is made within the framework of effective quantum mechanics; that is, the non-relativistic limit of a two-body scattering problem described by an underlying UV theory at a high energy scale is matched to a nonrelativistic quantum mechanics at a certain scale $\Lambda_{\rm QM}$, which is then used to describe scattering processes at lower energies. In scenarios where the Sommerfeld effect is relevant for the annihilation of a pair of particles, their interaction, namely, the potential $V(r)$ in the system's Hamiltonian, is typically separated into two parts:
\begin{align}
    \label{eq:potential}
    V(r) = V_L(r) + V_S(r),
\end{align}
where $V_L(r)$ is a long-range component representing their persistent interaction, and $V_S(r)$ is a short-range component responsible for inducing their annihilation. Schematically, the Sommerfeld effect arises due to the distortion of the plane wave associated with the annihilation via $V_S(r)$ caused by the long-range interaction $V_L(r)$. As a concrete example of those potentials encountered in actual problems, the long-range part is typically modeled by a Yukawa interaction, while the short-range part is assumed to consist of a delta function and its higher derivatives. For instance, in a system where a particle $\varphi$ of mass $M$ couples to a light vector mediator $A$ of mass $m_V$ with the structure constant $\alpha$, the dynamics of the interpolating field for the two-body state $\Psi(x,\vec{r}) \sim \varphi^\dagger(t,\vec{x}+\vec{r}/2) \varphi(t,\vec{x}-\vec{r}/2)$—after integrating out the hard scale $M$ and the soft scale $Mv$ with $v$ being the relative velocity—is described at the leading order by the following potential non-relativistic Lagrangian\,\cite{Pineda:1998kn, Brambilla:2004jw,Beneke:2022rjv,Biondini:2021ccr}:
\begin{align}
    \label{eq:pNR lagrangian}
    \mathcal{L}_{\rm pNR} \simeq \int d^3r \Psi^\dagger(x,\vec{r}) \left(i\partial_t + \frac{\nabla^2_{\vec{x}}}{4M} + \frac{\nabla_r^2}{M} + \alpha \frac{e^{-m_Vr}}{r} + i\frac{2\pi \alpha^2}{M^2} \delta(\vec{r}) \right) \Psi(x,\vec{r}) + \cdots.
\end{align}
From the Lagrangian (\ref{eq:pNR lagrangian}), one can move to the Hamiltonian formalism and, by projecting the Fock space spanned by two-body states onto the Hilbert space consisting of single two-body state, obtain a Hamiltonian with the potentials $V_L(r) = -\alpha e^{-m_Vr}/r$, $V_S(r) = -i2\pi \alpha^2 \delta(\vec{r})/M^2$.
The long-range and short-range potentials originate from the contributions of the soft and hard scales, respectively.

Let us describe the Sommerfeld effect as a scattering problem under the potential (\ref{eq:potential}). First, the Born expansion, which treats the potential as a mere perturbation, fails to yield a reliable result. In fact, under the first Born approximation with the Yukawa potential $V_L(r)=-\alpha e^{-m_Vr}/r$, the scattering amplitude behaves as $f(\vec{q}) = -2\alpha m/(\vec{q}^2+m^2)$ for the momentum transfer $\vec{q}$, and becomes singular in the non-relativistic limit $\vec{q}\to 0$ when the mass of the mediator $m$ is small. Moreover, due to the unitarity requirement, higher-order Born terms also exhibit similar singular behavior. Therefore, the Born approximation is not suitable in nonrelativistic regimes, and this is precisely the origin of the Sommerfeld effect.

To address this issue, one must therefore treat at least the long-range part of the potential $V_L(r)$ exactly, without approximation. For this purpose, it is useful to return to the Lippmann–Schwinger formalism. By using the Lippmann-Schwinger equation (\ref{eq:LS equation}) for the asymptotic state $\ket{\vec{p}^{-}_L}$ that incorporate only the effect of the long-range part of the potential, one can express the full scattering amplitude $f(\vec{p}\to \vec{p'}) = -(2\pi)^2 \mu \braket{\vec{p'}|V|\vec{p}^{+}}$ in the form of the so-called two-potential formula\,\cite{Weinberg:1995mt,Taylor:1972pty}:
\begin{align}
    \label{eq:two-potential formalism}
    f(\vec{p} \to \vec{p'}) = f^L(\vec{p} \to \vec{p'}) -(2\pi)^2\mu \braket{\vec{p}^{'-}_L|V_S|\vec{p}^+},
\end{align}
where $f^L(\vec{p}\to \vec{p'}) = -(2\pi)^2 \mu \braket{\vec{p'}|V_L|\vec{p}_L^{+}}$ is the scattering amplitude which includes only the effect of the long-range potential $V_L$. The relation (\ref{eq:two-potential formalism}) separates the effect of the short-range part from the contribution involving only the long-range interaction, making it suitable for a perturbative expansion in $V_S$. By expanding the asymptotic state $\ket{\vec{p}^+}$ in the second term around those for the long-range potential, $\ket{\vec{p}_L^+}$, and further performing the partial-wave expansion one obtains the distorted-wave Born approximation (DWBA) for the partial wave amplitude:
\begin{align}
    \label{eq:DWBA}
    f_{\ell}(p) \simeq f^L_\ell(p) - \frac{2\mu}{p^2} \int^\infty_0 dr \,\psi^L_{\ell,p}(r) V_S(r) \psi^L_{\ell,p}(r).
\end{align}
Here we denote the $\ell$-th partial-wave amplitude and the scattering wave function associated with the long-range potential $V_L(r)$ as $f_\ell^L(p)$ and $\psi^L_{\ell,p}(r)$, respectively. Recalling that the wave function $\psi^L_{\ell,p}(r)$ can be expressed in terms of the regular solution $\phi_{\ell,p}(r)$ and the Jost function $\mathscr{J}_\ell(p)$ of the potential as $\psi^{L}_{\ell,p}(r)=\phi_{\ell,p}(r)/\mathscr{J}_\ell(p)$, and that the short-range potential $V_S(r)$ behaves like a delta function and its derivatives and thus the regular solution is replaced by the Riccati-Bessel function $j_\ell(pr)$ near the origin, the expression (\ref{eq:DWBA}) can be simplified as follows:
\begin{align}
    \label{eq:DWBA simplified}
    f_\ell(p) \simeq f^L_\ell(p) + \frac{1}{\mathscr{J}_\ell(p)^2}f^S_\ell(p), \quad
    f^S_\ell(p) \simeq -\frac{2\mu}{p^2}\int^\infty_0 dr \, j_\ell(pr)V_S(r)j_\ell(pr).
\end{align}
Note that the $f^S_\ell(p)$ appearing in the above equation is nothing but the Born amplitude due to the short-range potential. In fact, the explicit form of the potential $V_S(r)$ is determined by the matching procedure requiring that the above expression reproduces the non-relativistic limit of the amplitude derived from the underlying UV theory. This observation indicates that, for practical purposes, the explicit form of $V_S(r)$ is unnecessary, as long as it is short-ranged. Therefore, it is sufficient to adopt the non-relativistic limit of the UV amplitude as $f^S_\ell(p)$.

We are now in a position to compute the Sommerfeld enhancement induced by the long-range potential using the optical theorem. The annihilation cross section for the $\ell$-th partial wave, $(\sigma_\ell v)^{\rm ann}$, is given by the difference between the total and elastic cross sections, $(\sigma_\ell v)^{\rm tot}$ and $(\sigma_\ell v)^{\rm el} $, with the total cross section given by the optical theorem: 
\begin{align}
\label{eq:optical annhilation}
    (\sigma_\ell v)^{\rm tot} = (\sigma_\ell v)^{\rm el}+ (\sigma_\ell v)^{\rm ann}, \quad (\sigma_\ell v)^{\rm tot}  = \frac{4\pi(2\ell+1)}{\mu}{\rm Im}\,f_\ell(p), \quad (\sigma_\ell v)^{\rm el}  = \frac{4\pi(2\ell+1)}{\mu}p|f_\ell(p)|^2.
\end{align}
Assuming, for simplicity, that the contribution from the real part of the UV amplitude $f_\ell^S(p)$ is negligible, and its imaginary part—determined by the optical theorem—gives the tree-level annihilation cross section, $(\sigma_\ell v)^{\rm ann,0}$, we can find the approximation $f^S_\ell(p) \simeq i\mu (\sigma_\ell v)^{\rm ann,0} / 4\pi(2\ell+1)$, from which Eq.~(\ref{eq:DWBA simplified}) yields the following factorized expression for the annihilation cross section including the Sommerfeld effect:
\begin{align}
    \label{eq:def SE}
    (\sigma_\ell v)^{\rm ann}  \simeq S_\ell(p) (\sigma_\ell v)^{\rm ann,0}, \quad S_\ell(p) = \frac{1}{|\mathscr{J}_\ell(p)|^2}.
\end{align}
Here, in the above expression, we have retained only the leading term in $f^S_\ell(p)$ and used the relation $\mathscr{J}_\ell(p) = |\mathscr{J}_\ell(p)|e^{-i\delta_\ell(p)}$ for $p>0$. We see from Eq.~(\ref{eq:def SE}) that the Sommerfeld factor for the $\ell$-th partial wave $S_\ell(p)$ is expressed in terms of the Jost function, which encodes the spectral properties of the quantum system under the long-range potential $V_L(r)$.

This fact clearly indicates that the Sommerfeld factor exhibits a sharp peak at the corresponding momentum whenever the system possesses special energy eigenstates.
One possible situation of this kind is when the system possesses a shallow bound state under the long-range potential. As discussed in the previous section, this corresponds to the Jost function having a zero near the origin on the positive imaginary axis. Consequently, the corresponding Sommerfeld factor exhibits a peak at small positive values of $p$, induced by this bound state. In particular, let us consider the behavior in the presence of a zero-energy resonance, namely when the Jost function has a zero at the origin. In this case, Eq.~(\ref{eq:low energy expansion of jost}) implies that the Jost function can be expanded around $p=0$ as$\mathscr{J}_\ell(p) = \beta_\ell p^2 + i\gamma_\ell p^{2\ell+1} + \mathcal{O}(p^4)$, and thus, on the threshold, the Sommerfeld factor behaves as\,\cite{Kamada:2023iol}
\begin{align}
    \label{eq:threshold scaling}
    S_\ell(p) \propto
    \left
    \{\begin{aligned}
    p^{-2} \quad (\ell = 0)\\
    p^{-4} \quad (\ell \ge 1)
    \end{aligned}
    \right. .
\end{align}
Another possibility is the existence of a resonance state in the spectrum, namely when the Jost function has a zero in the region ${\rm Re}\,p>0$\,\cite{Beneke:2024iev}. In this case, if the zero lies not too far from the real axis, that is, the decay width of the resonance is sufficiently small, one expects the Sommerfeld factor to exhibit a resonance feature when the energy of scattering particles approaches the resonance energy, much like how the elastic scattering cross section displays a Breit–Wigner–type resonance. However, we must keep in mind that the actual annihilation cross section as a function of momentum $p$ may not exhibit such resonance behavior. This is because, in deriving Eq.~(\ref{eq:def SE}), we have neglected the $\mathcal{O}(\mathscr{J}_\ell(p)^{-4})$ term arising from the square of the UV amplitude, and this approximation is no longer valid in the momentum region close to the corresponding resonance energy. This observation naturally leads to the subject of the next section, which concerns the unitarization of the Sommerfeld effect.

The derivation of the Sommerfeld factor presented here is not entirely standard; in fact, it is often introduced differently in the literature. For example, in\,\cite{Cassel:2009wt,Iengo:2009ni}, the Sommerfeld factor for the $\ell$-th partial wave $S_\ell(p)$ is given, in our notation, by
\begin{align}
    S_\ell(p) = \left |\frac{(2\ell +1)!!}{ p^{\ell +1} (\ell+1)!} \left(\frac{\partial^{\ell+1}}{\partial r^{\ell+1}}\psi^L_{\ell,p}(r) \right )_{r=0} \right  |^2
\end{align}
This expression is obviously equivalent to our result in Eq.~(\ref{eq:def SE}). The factorial factors and derivative operations appearing in the above formula simply correspond to the procedure of extracting the Jost function from the scattering wave function $\psi^L_{\ell,p}(r)$. The reason we have introduced the Sommerfeld factor in a manner that differs from the conventional approach is to highlight two key aspects: first, the way in which its momentum dependence reflects the spectral properties of the system, and second, the fact that the Sommerfeld factor itself is derived based on certain approximations.

Finally, although, in the above discussion, we have used the optical theorem to compute the annihilation cross section and the Sommerfeld factor, at leading order the annihilation effect can also be incorporated directly at the amplitude level by employing the unitarity relation (\ref{eq:optical theorem}) for coupled channels. Labeling the channel of parent particles as 1 and the channel of daughter particles as 2, the unitarity relation (\ref{eq:optical theorem}) can be written explicitly as:
\begin{align}
    \label{eq:unitarity relation}
    {\rm Im}\, f_{\ell\,11}(p) \simeq p|f_{\ell\,11}(p)|^2, \quad {\rm Im}\, f_{\ell\,12} (p)\simeq p f_{\ell\,11} (p)f^*_{\ell\,12}(p),
\end{align}
where $f_{\ell\,11}(p)$ and $f_{\ell\,12}(p)$ are the $\ell$-th partial-wave amplitude for the self-scattering $1\to1$ and the pair annihilation $1 \to 2$ respectively, and we neglected the contribution of the unitarity cut on the RHS of the optical theorem associated with the threshold of the daughter particles, assuming the hierarchy $f_{\ell\,11} \gg f_{\ell\,12} \gg f_{\ell\,22}$. While the first equation in $(\ref{eq:unitarity relation})$ corresponds to the standard unitarity relation for a single-channel scattering amplitude, the second equation is particularly notable: it implies that the phases of $f_{\ell\,11}$ and $f_{\ell\,12}$ are identical on the unitarity cut of the parent particles. This truncated form of the unitarity relation is known as Watson's theorem\,\cite{Kamada:2023iol,ParticleDataGroup:2024cfk,Oller:2020guq}. The functional form of $f_{\ell\,12}(p)$ that satisfies Watson’s theorem can be readily determined in the present case by making use of the Jost function associated with $f_{\ell\,11}(p)$. By assigning the discontinuity associated with the parent particle threshold to the Jost function, we can write $f_{\ell\,12}(p) = \mathcal{M}_\ell(p)/\mathscr{J}_\ell(p)$, where $\mathcal{M}_\ell(p)$ is a function without the branch cuts. One can easily verify that this expression indeed satisfies Watson’s theorem by using the relation ${\rm Im}\, \mathscr{J}_\ell(p) = -p \mathscr{J}_\ell(p) f_{\ell\,11}(p)$. The undetermined function $\mathcal{M}_\ell(p)$ must encode the short-distance annihilation dynamics and is free of the singularities introduced by the long-range potential, which are entirely captured by the Jost function $\mathscr{J}_\ell(p)$. We therefore match $\mathcal{M}_\ell(p)$ to the tree-level annihilation amplitude $f^{\rm tree}_{\ell\,12}(p)$, yielding the full annihilation amplitude including the long-range effect as\footnote{Ref.\cite{Kamada:2023iol} investigates the Sommerfeld enhancement starting from Watson’s theorem, and explores its relation to the self-scattering amplitude by employing Levinson’s theorem and a dispersion relation. In contrast, our approach introduces the Jost function from the outset and directly solves Watson’s theorem in terms of it. Within this framework, both Levinson’s theorem and the dispersion relation emerge as consequences of the analytic properties of the Jost function. In particular, the correlation between the Sommerfeld enhancement factor and the self-scattering amplitude is encoded in the Hilbert transform relating the modulus and phase of the Jost function. See Appendix~\ref{app: Jost} for further details.}
\begin{align}
    f_{\ell\,12}(p) \simeq f^{\rm tree}_{\ell\,12}(p)/\mathscr{J}_\ell(p).
\end{align}
This separation allows us to incorporate the Sommerfeld enhancement effects into the annihilation amplitude itself straightforwardly, as all the nontrivial energy dependence due to the long-range force is contained in the denominator. However, as mentioned before, in the region where the enhancement due to the Jost function becomes large, the effect of the cut associated with the daughter particles can no longer be neglected, and thus the above expression loses its validity.

\section{Improvement of the scattering amplitude}
\label{sec: RG improve}

As discussed in the previous section, when the system possesses a shallow bound state, the conventional Sommerfeld factor $S_\ell(p)$ in Eq.~(\ref{eq:def SE}) exhibits a strong peak at small momenta due to a factor originating from the Jost function. In particular, near a zero-energy resonance, it scales according to Eq.~(\ref{eq:threshold scaling}). Combining this with the typical dependence of the tree-level annihilation cross section for the $\ell$-th partial wave, $(\sigma_\ell v)^{\rm ann,0}\propto p^{2\ell}$, the long-range enhanced annihilation cross section scales as
\begin{align}
    \label{eq:scale danger}
    (\sigma_\ell v)^{\rm ann}
    \propto 
    \left \{
    \begin{aligned}
    p^{-2} \quad (\ell = 0)\\
    p^{2\ell-4} \quad (\ell \ge 1)
    \end{aligned}
    \right. ,
\end{align}
which shows a singular behavior in $p\to0$ for the $s$-wave ($\ell=0$) and the $p$-wave ($\ell=1$).

On the other hand, the optical theorem Eq.~(\ref{eq:optical annhilation}) allows the annihilation cross section to be expressed as (with $s_\ell$ denoting the $S$-matrix element for self-scattering in the $\ell$-th partial wave)
\begin{align}
(\sigma_\ell v)^{\rm ann} = \frac{4\pi(2\ell+1)}{\mu p} \frac{1 - |s_\ell|^2}{4},
\end{align}
from which, together with the unitarity condition $|s_\ell| \leq 1$, one obtains the following unitarity bound that provides a universal upper limit on the annihilation cross section:
\begin{align}
\label{eq:unitarity bound}
(\sigma_\ell v)^{\rm ann} \leq \frac{\pi(2\ell+1)}{\mu p}.
\end{align}
Comparing this with Eq.~(\ref{eq:scale danger}), we find that the scaling behavior just above a zero-energy resonance exhibits a stronger singularity for $\ell=0,1$ than what is permitted by unitarity. This indicates that the naive expression for the Sommerfeld factor Eq.~(\ref{eq:def SE}) can lead to a violation of unitarity in the presence of a shallow bound state, and its use in such cases is not justified.

We discuss the restoration of unitarity in the Sommerfeld effect below. We first identify the cause of the violation as originating from the fact that the DWBA, which plays a central role in the analysis of the Sommerfeld effect, constitutes a singular perturbation. As a result, its convergence becomes delicate at small $p$ in the presence of shallow bound states. We then introduce the renormalization group technique to resum and improve the naive perturbative series. Finally, we examine the properties of the improved Sommerfeld effect, and demonstrate its validity by applying it to the specific case of a spherical well potential.

\subsection{Origin of the unitarity violation}
\label{Cause of unitarity violation}

The origin of the unitarity violation in the Sommerfeld enhancement (\ref{eq:def SE}) can be traced back to its derivation of Eq.~(\ref{eq:def SE}) in the previous section. In that analysis, we applied the optical theorem to the self-scattering amplitude (\ref{eq:DWBA simplified}) at the lowest order in DWBA, retaining only the leading-order contribution from the short-range interaction. However, as previously noted, in the presence of a shallow bound state, the perturbation terms, which involve inverse powers of the Jost function, become singular in the limit $p \to 0$. Consequently, higher-order contributions such as the second term in Eq.~(\ref{eq:DWBA simplified}), which would be subleading in perturbation theory, can dominate over the leading-order term. In other words, the perturbative expansion (\ref{eq:DWBA simplified}) has secular terms under the existence of shallow bound states and becomes a singular perturbation. It is important to note that this behavior is independent of the strength of the short-range potential; the infrared divergence arises solely from the properties of the long-range potential.

The appearance of singular perturbation theory is not uncommon in physics. As a simple example from particle physics, consider the $\phi^4$ theory. In the one-loop approximation, the renormalized four-point amplitude at renormalization scale $\mu$ takes the form
\begin{align}
    \label{eq:phi-4}
    \mathcal{M}(p^2) = \lambda(\mu) + \frac{3\lambda^2(\mu)}{32\pi^2} \log\left( \frac{p^2 + m^2}{\mu^2} \right) + \cdots ,
\end{align}
where $\lambda(\mu)$ is the renormalized coupling constant at the scale $\mu$ and $m$ is the mass parameter. When the momentum scale is significantly different from the renormalization scale $\mu$, the logarithmic correction becomes large, thereby invalidating the naive perturbative expansion. Note that this breakdown occurs regardless of the strength of the coupling constant. In fact, the logarithmic terms that appear in higher orders of perturbation are essential for reproducing the branch cut associated with the low-energy particle threshold in the scattering amplitude. 

Therefore, the naive perturbative expansion~(\ref{eq:phi-4}) is not sufficient, and we must improve its convergence by resumming the secular terms arising from the large logarithms. Fortunately, we know how to achieve this for~(\ref{eq:phi-4}): we apply the method of the renormalization group (RG). By requiring that the right-hand side does not depend on the artificial renormalization scale $\mu$, we derive the renormalization group equation, whose solution effectively resums the contributions from a specific class of diagrams. This combined approach of the naive perturbative expansion and the renormalization group is known as renormalized perturbation theory.

This idea of resummation via the RG suggests a potential resolution to the problem of singular perturbation in the DWBA, since the two perturbative expansions—Eq.~(\ref{eq:phi-4}) and Eq.~(\ref{eq:DWBA simplified})—share a parallel structure: both contain secular terms originating from the infrared properties of the system, and their presence is independent of the magnitude of the coupling constant that counts the order of perturbation. Indeed, it is well known that singular perturbation problems can often be addressed using RG techniques\,\cite{Chen:1994zza,Chen:1995ena}. In this paper, we propose to improve Eq.~(\ref{eq:DWBA simplified}) by applying the RG method to systematically resum the higher-order contributions that give rise to secular terms. This procedure effectively transforms the naive perturbative expansion~(\ref{eq:DWBA simplified}) into a renormalized perturbation theory, thereby restoring its validity and extending its applicability.

The necessity of resummation for unitarization is further motivated from a more physical perspective. As an elementary example, let us recall the scattering amplitude arising from an $s$-channel diagram. In an $s$-channel resonance at leading order, the scattering amplitude diverges when the energy approaches the mass of the mediator, as the propagator hits its pole, and this leads to a violation of the unitarity bound. In reality, such singular behavior is smeared out by the width of the mediator, which arises from resumming its self-energy corrections. In other words, to avoid the divergence, it is essential to consistently resum higher-order perturbative contributions.

An analogous situation is expected for the Sommerfeld enhancement. Let us therefore consider what the actual unitarized behavior of the Sommerfeld enhancement should be, in light of the mechanisms that regulate singularities in simpler contexts such as the $s$-channel resonance.\footnote{ In the presence of a shallow bound state, if we describe the self-scattering amplitude using an effective field theory, we must introduce not only fields for the scattering particles but also an additional field corresponding to the bound state that couples to them. In this framework, the self-scattering process is described as an $s$-channel resonance mediated by the bound-state field.}For example, in the zero-range approximation\,\cite{Braaten:2013tza} of the $s$-wave elastic scattering amplitude in the absence of annihilation, namely considering only a positive scattering length $a_0>0$ in Eq.~(\ref{eq:effective range expansion}), the self-scattering amplitude is given by
\begin{align}
    f_0(p) = \frac{1}{-1/a_0 - ip},
\end{align}
which has a bound state pole at $p_B=i/a_0$.
When the annihilation effect is turned on, the scattering length acquires an imaginary part, and the bound-state pole shifts off the imaginary axis, signaling its instability. Specifically, if the inverse scattering length shifts as $-1/a \to -1/a -i\gamma$ with $\gamma>0$\footnote{The positivity of the decay width of the bound state $\Gamma = 2\gamma/\mu a_0$ requires that the sign of $\gamma$ be positive.}, the optical theorem implies that the annihilation cross section becomes\,\cite{Braaten:2013tza,Braaten:2017gpq,Chu:2019awd}
\begin{align}
    (\sigma_0 v)^{\rm ann} = \frac{4\pi}{\mu} \frac{\gamma}{|-1/a_0-i(p+\gamma)|^2}.
\end{align}
This expression is manifestly consistent with the unitarity bound. In fact, when the scattering length is large—corresponding to a shallow bound state—the imaginary part in the denominator regulates the excessive enhancement at small $p$. In particular, even in the limit of a zero-energy resonance, where the scattering length diverges, the annihilation cross section remains finite and constant. Accordingly, by analogy with the $s$-channel resonance case, the introduction of a decay width for the bound state necessitates an appropriate resummation of the perturbative series.

\subsection{An illustrative example of the RG method}

We now introduce the RG method for singular perturbations. Before addressing the improvement of the DWBA~(\ref{eq:DWBA simplified}), we first illustrate the RG method through a simple example from elementary calculus. Let us consider the following ordinary differential equation:
\begin{align}
    \label{eq:differential equation}
    \frac{dy}{dt} - y = \varepsilon y^3,
\end{align}
where $\varepsilon$ is an infinitesimal parameter treated as a perturbation. The naive perturbative solution to the equation is then given, up to the first-order correction, by
\begin{align}
    \label{eq:RG example naive}
    y(t) \simeq Ae^t + \frac{1}{2}\varepsilon A^3e^{3t} + \mathcal{O}(\varepsilon^2).
\end{align}
Here, the integration constant is denoted by $A$. In the above expression, the second term contains not only the perturbation parameter $\varepsilon$, but also a factor of $A^3e^{3t}$ originating from the leading-order term $Ae 
^t$. This constitutes a secular term, and as a result, the perturbative correction eventually dominates the unperturbed part for sufficiently large $t$, rendering the perturbative expansion invalid in that regime. To improve the situation, we introduce a trivial factor $CZ=1$ into the unperturbed solution, which serves to make the dependence on the unperturbed part for each term explicit:
\begin{align}
    \label{eq:differential equation2}
    y(t) \simeq CZAe^t + \frac{1}{2}\varepsilon (CZ)^3A^3e^{3t} + \mathcal{O}(\varepsilon^2).
\end{align}
We allow $C$ and $Z$ to depend on an arbitrary parameter $\tau$, and expand $Z$ in powers of $\varepsilon$, as $Z=1+\varepsilon Z_1 + \cdots$. Then Eq.~(\ref{eq:differential equation2}) becomes, up to the order of $\varepsilon$,
\begin{align}
    y \simeq AC(\tau)e^t + \varepsilon AC(\tau)\left(\frac{1}{2}A^2C(\tau)^2e^{2t} + Z_1(\tau) \right) + \mathcal{O}(\varepsilon^2). 
\end{align}
We now fix the $\tau$-dependence of $Z_1$ such that it cancels the secular term at $t=\tau$, as $Z_1(\tau) = -A^2C(\tau)^2e^{2\tau}/2$. The resulting improved solution is then given by
\begin{align}
    \label{eq:RG example solution}
    y^{\rm imp}(t) \simeq AC(t)e^t.
\end{align}
The secular term is absorbed into the counterterm $Z(\tau)$, and the improved solution takes the form of the unperturbed part multiplied by a "renormalized integration constant" $C(\tau)$. The $\tau$-dependence of $C(\tau)$ is determined by requiring that the solution does not depend on the artificial parameter $\tau$. From this, the renormalization group equation up to first order in the perturbation is given by
\begin{align}
    \label{eq: RG example}
    \frac{dC(\tau)}{d\tau} \simeq \varepsilon C(\tau)^3A^2 e^{2\tau}
\end{align}
By integrating the renormalization group equation from $\tau_0$ to $\tau$ and combining it with Eq.~(\ref{eq:RG example solution}), we obtain the following result:
\begin{align}
    \label{eq:RG example improved}
    y^{\rm imp}(t) \simeq \pm \frac{Ae^t}{\sqrt{C^{-2}(\tau_0)-\varepsilon A^2 (e^{2\tau}-e^{2\tau_0})}}
\end{align}
As can be readily verified, expanding the improved solution above in powers of $\varepsilon$ reproduces the original naive perturbative series Eq.~(\ref{eq:RG example naive}). However, this expansion also contains an infinite series of secular terms of the form $\varepsilon^n e^{(2n+1)t}$, which implies that Eq.~(\ref{eq:RG example solution}) effectively resums the leading-order secular divergences that first appear at order $\varepsilon$.\footnote{In fact, under suitable initial conditions, the improved solution coincides with the exact solution of the original differential equation. This is because in the present case, all secular terms are of the form $\varepsilon^n e^{(2n+1)t}$, and the renormalization group equation becomes exact at leading order in $\varepsilon$.}

\subsection{Unitarization of Sommerfeld enhancement via the RG}

\label{sec:RG Sommerfeld}

We apply the RG method to improve the DWBA (\ref{eq:DWBA simplified}). The computational procedure we follow here is similar to that commonly employed in effective field theory\,\cite{Manohar:2018aog,Henning:2014wua}: we assume the existence of some UV theory at sufficiently high energy scales, and match it onto an effective quantum mechanics at low energies, in which the potential is given by (\ref{eq:potential}). To extract physical observables from the resulting effective theory, we perform renormalized perturbation theory, thereby avoiding secular terms arising from infrared singularities.

 The basic strategy for the renormalized version of the DWBA (\ref{eq:DWBA simplified}) is the same as in the previous differential equation example: introduce a counterterm into the naive perturbative expansion, use it to cancel the secular terms, derive the RG equation, and resum them by solving it. To make the dependence on the unperturbed contribution explicit, we introduce a trivial identity $CZ = 1$ into the naive expansion of the partial-wave scattering amplitude in Eq.~(\ref{eq:DWBA simplified}). Since the first DWBA correction term is determined by the integral $\int \psi^L V_S \psi^L$,
which scales quadratically with the leading-order wavefunction $\psi^L$, it is natural to assign a factor of $(CZ)^2$ to it. This leads to the modified expansion:
\begin{align}
    f_\ell(p) \simeq C(Q)f_\ell^L(p) + C(Q)\left(\frac{C(Q)}{\mathscr{J}_\ell(p)^2}f_\ell^S(p) + Z_1(Q)f_\ell^L(p)\right),
\end{align}
where $C(Q)$ and $Z(Q)$ are introduced as functions of an arbitrary momentum scale $Q$, and we expand as $Z(Q) = 1 + Z_1(Q) + \cdots$ with respect to the short range parameter contained in $V_S$.
We then choose the $Q$-dependence of the counterterm $Z(Q)$ such that it cancels the secular term at $p = Q$, leading to the RG-improved amplitude and the explicit form of the counterterm:
\begin{align}
    \label{eq:RG improved}
    f_\ell^{\rm imp}(p) \simeq C(p)f_\ell^L(p),\quad Z(Q) \simeq 1-\frac{C(Q)}{\mathscr{J}_\ell(Q)^2}\frac{f^S_\ell(Q)}{f^L_\ell(Q)}.
\end{align}
Note that the second term of the counterterm includes not only the UV amplitude $f_\ell^S(Q)$, which is proportional to the short-range parameter in $V_S$ and assumed to be small, but also the inverse of the long-range amplitude $f^L_\ell(Q)$. This means that, for the perturbative renormalization applied here to work properly, there must exist a hierarchy between the long-range and short-range contributions to the scattering amplitude $f_\ell(p)$ at some renormalization scale $Q$. 

Demanding that the full amplitude be independent of the arbitrary matching scale $Q$ yields the following renormalization group equation, which determine the renormalization scale dependence of $C(Q)$:
\begin{align}
    \frac{dC(Q)}{dQ} \simeq C(Q)^2\frac{d}{dQ}\left(\frac{1}{\mathscr{J}_\ell(Q)}\frac{f_\ell^S(Q)}{f_\ell^L(Q)}\right).
\end{align}
Integrating this equation from some matching scale $p_0$ and substituting back into the amplitude expression in Eq.~(\ref{eq:RG improved}) gives the RG-improved form of the self-scattering amplitude:
\begin{align}
    \label{eq:RG improved amplitude}
    f^{\rm imp}_\ell(p) \simeq \frac{f_\ell^L(p)}{C(p_0)^{-1}-\frac{1}{\mathscr{J}_\ell(p)^2}\frac{f_\ell^S(p)}{f_\ell^L(p)}+\frac{1}{\mathscr{J}_\ell(p_0)^2}\frac{f_\ell^S(p_0)}{f_\ell^L(p_0)}}.
\end{align}
The matching scale $p_0$ must be chosen within the regime where the effective quantum mechanical picture remains reliable. Since the effective description is only valid below the cutoff scale $\Lambda_{\rm QM}$—defined as the matching point with the underlying UV theory—$p_0$ must lie below $\Lambda_{\rm QM}$. Here, we set $p_0$ at (or near) the cutoff, $p_0 \simeq \Lambda_{\rm QM}$, where the amplitude is well approximated by the naive DWBA expression Eq.~(\ref{eq:DWBA simplified}), and then the constant $C(p_0)$ is fixed to be
\begin{align}
    \label{eq:RG initial constant}
    C(p_0) = 1 + \frac{1}{\mathscr{J}_\ell(p_0)}\frac{f_\ell^S(p_0)}{f_\ell^L(p_0)}.
\end{align}
This matching procedure ensures that the correction term remains small, so the second term in Eq.~(\ref{eq:RG initial constant}) can be treated perturbatively. Under this approximation, the $p_0$-dependence of the denominator in Eq.~(\ref{eq:RG improved amplitude}) cancels and the improved amplitude simplifies to the following form:
\begin{align}
    \label{eq:RG improved amplitude simplified}
    f^{\rm imp}_\ell(p) \simeq \frac{f_\ell^L(p)}{1-\frac{1}{\mathscr{J}_\ell(p)^2}\frac{f_\ell^S(p)}{f_\ell^L(p)}} = \frac{\mathscr{J}_\ell(-p)-\mathscr{J}_\ell(p)}{2ip\left(\mathscr{J}_\ell(p)-\frac{2ipf_\ell^S(p)}{\mathscr{J}_\ell(-p)-\mathscr{J}_\ell(p)}\right)},
\end{align}
where we used Eq.~(\ref{eq:analytic s matrix}) in the second equality.
This expression constitutes the RG-improved scattering amplitude that resums the leading secular behavior arising in the DWBA under the influence of long-range potentials, and is valid even in the presence of shallow bound states.
Note that the expression Eq.~(\ref{eq:RG improved amplitude simplified}) allows us to interpret the improvement as an effective shift of the Jost function:
\begin{align}
    \label{eq:improved jost}
    \mathscr{J}^{\rm imp}_\ell(p) = \mathscr{J}_\ell(p) - \frac{2ip f_\ell^S(p)}{\mathscr{J}_\ell(-p)-\mathscr{J}_\ell(p)}, \quad f^{\rm imp}_\ell(p) = \frac{\mathscr{J}_\ell^{\rm imp}(-p)-\mathscr{J}_\ell^{\rm imp}(p)}{2ip\mathscr{J_\ell^{\rm imp}}(p)},
\end{align}
where we used that the UV amplitude $f_\ell^{S}(p)$ is a even function of $p$ and thus $f_\ell^{S}(p)=f^{S}_\ell(-p)$. 

Let us remark that the simplified version of the amplitude Eq.~(\ref{eq:improved jost}) obtained using the RG method can also be derived on the basis of the following, more intuitive argument. In general, solving the Schr\"{o}dinger equation that includes a highly singular short-range potential—such as one given by a delta function and its derivatives—is not straightforward (see \cite{Blum:2016nrz,Parikh:2024mwa} and the discussion in Sec~\ref{sec: Wilson}). Suppose, however, that it can be solved in some sense (for example, under an appropriate regularization and matching). Then, as in ordinary potential scattering, the Schr\"{o}dinger equation admits scattering solution, $\psi_{\ell,p}^{\rm mod}(r)$, and its behavior near the origin should be expressed in terms of a Jost function $\mathscr{J}_\ell^{\rm mod}(p)$ that is modified to incorporate the effects of the singular potential, $\psi^{\rm mod}_{\ell,p}(r) \to j_\ell(pr)/\mathscr{J}_\ell^{\rm mod}(p)$. Combining this with the two-potential formula Eq.~(\ref{eq:two-potential formalism}), the modified self-scattering amplitude $f_\ell^{\rm mod}(p)$ including the effects of the short-range interaction becomes
\begin{align}
f_\ell^{\rm mod}(p) = f^L_\ell(p) + \frac{1}{\mathscr{J}_\ell(p)}\frac{1}{\mathscr{J}_\ell^{\rm mod}(p)}f_\ell^S(p).
\end{align}
On the other hand, taking into account the role of the Jost function in scattering theory, the modified self-scattering amplitude should be written using the modified Jost function as 
\begin{align}
f_\ell^{\rm mod}(p) = \frac{\mathscr{J}^{\rm mod}_\ell(-p)-\mathscr{J}_\ell^{\rm mod}(p)}{2ip\mathscr{J}^{\rm mod}_\ell(p)}.
\end{align}
Only the part of the modified Jost function that is odd in $p$ contributes to the numerator above. From Eq.~(\ref{eq:low energy expansion of jost}), the source of the $p$-odd piece is the unitarity cut arising from self-scattering due to the long-range potential. Therefore, as a first approximation, we may take all contributions of the short-range potential—including annihilation effects—to the modified Jost function to be $p$-even, and thus the $\mathscr{J}^{\rm mod}_\ell(p)$ in the numerator can be replaced by the Jost function $\mathscr{J}_\ell(p)$ due solely to the long-range potential. Then we find the self-consistent equation for the modified Jost function
\begin{align}
 \frac{\mathscr{J}_\ell(-p)-\mathscr{J}_\ell(p)}{2ip\mathscr{J}^{\rm mod}_\ell(p)}= f^L_\ell(p) + \frac{1}{\mathscr{J}_\ell(p)}\frac{1}{\mathscr{J}_\ell^{\rm mod}(p)}f_\ell^S(p),
\end{align}
whose solution leads us the improved Jost function in Eq.~(\ref{eq:improved jost}). This line of reasoning may also provide an intuitive understanding of when Eq.~(\ref{eq:improved jost}) obtained from the RG method breaks down: if the contributions of the long-range and short-range potentials to the scattering become comparable, the effect of the unitarity cut can no longer be represented by the long-range Jost function alone, and the numerator-replacement trick ceases to be applicable.

It is now straightforward to obtain the annihilation cross section incorporating the improved Sommerfeld enhancement. One simply applies the relation Eq.~(\ref{eq:RG improved amplitude}) and Eq.~(\ref{eq:RG initial constant}) or Eq.~(\ref{eq:RG improved amplitude simplified}) to the optical theorem Eq.~(\ref{eq:optical annhilation}). For simplicity, we adopt the simplified version Eq.~(\ref{eq:RG improved amplitude simplified}) and the approximation for the UV amplitude $f^S_\ell(p) \simeq i\mu (\sigma_\ell v)^{\rm ann,0} / 4\pi(2\ell+1)$, and obtain the improved Sommerfeld factor $S_\ell^{\rm imp}(p)$ as
\begin{align}
    \label{eq:improved SE}
    (\sigma_\ell v)^{\rm ann} = S_\ell^{\rm imp}(p) (\sigma_\ell v)^{\rm ann,0}, \quad S_{\ell}^{\rm imp}(p) = \frac{1}{|\mathscr{J}^{\rm imp}_\ell(p)|^2}.
\end{align}
Eq.~(\ref{eq:improved SE}) represents the expression for the unitarized Sommerfeld factor proposed in this paper. Compared to the conventional Sommerfeld factor in Eq.~(\ref{eq:def SE}), $S_\ell(p)=1/|\mathscr{J}_\ell(p)|^2$, the Jost function has been replaced by its improved counterpart $\mathscr{J}^{\rm imp}_\ell(p)$. Therefore, using the explicit form of the improved Jost function Eq.~(\ref{eq:improved jost}) and the fact that the Jost function is given by $\mathscr{J}_\ell(p) = |\mathscr{J}_\ell(p)|e^{-i\delta_\ell(p)}$ for $p > 0$, we find that unitarization requires not only the conventional Sommerfeld factor, but also additional input: the UV amplitude $f_\ell^S(p)$ and the phase shift $\delta_\ell(p)$.

To conclude this subsection, we examine the range of validity of the improved amplitude in Eq.~(\ref{eq:RG improved amplitude simplified}). Its applicability is subject to two important caveats for the renormalization procedure arising in its derivation.
(i) The effective quantum mechanical framework is applicable only below a certain cutoff scale $\Lambda_{\rm QM}$, which defines the boundary of the low-energy regime. Consequently, Eq.~(\ref{eq:RG improved amplitude simplified}) should not be applied at momenta exceeding this cutoff, $p > \Lambda_{\rm QM}$. (ii) The contribution to the self-scattering amplitude from the short-range potential must be sufficiently smaller than that from the long-range potential. This hierarchy is essential for justifying the perturbative treatment of the counterterm and the cancellation of the secular term. Two typical situations in which this condition is violated are: (a) the long-range potential is too weak to dominate the scattering, or (b) the short-range potential is so strong that it must be treated non-perturbatively. However, case (a) corresponds to the situation where the Born approximation is already sufficient, making the improvement unnecessary. In this regime, the Sommerfeld enhancement itself is not expected to play a significant role. Case (b) will be commented on in Sec~\ref{sec: Wilson}.

\subsection{Behavior of the improved scattering amplitude}

In this subsection, we investigate the properties of the improved amplitude Eq.~(\ref{eq:improved jost}) and the improved annihilation cross section Eq.~(\ref{eq:improved SE}). More specifically, we discuss their consistency with the unitarity bound Eq.~(\ref{eq:unitarity bound}), their behavior in low-energy scattering, and their poles. As an illustrative example, we also examine the application of the RG improvement to the scattering problem with a square-well potential, for which the Jost function is analytically tractable.

Let us begin by verifying that the improved annihilation cross section satisfies the unitarity bound. The decomposition of the Jost function for the long-range potential given in Eq.~(\ref{eq:low energy expansion of jost}) implies that the real part of the improved Jost function $\mathscr{J}^{\rm imp}_\ell(p)$ in Eq.~(\ref{eq:improved jost}) is an even function of $p$, while its imaginary part is given by
\begin{align}
    {\rm Im} \mathscr{J}_\ell^{\rm imp}(p) = p^{2\ell+1} G_\ell(p^2) \left[1+\frac{{\rm Im}f_\ell^S(p)}{p^{4\ell+1}G^2_\ell(p^2)}\right],
\end{align}
from which, using the fact that ${\rm Im} f_\ell^S(p) > 0$, we obtain the inequality $({\rm Im} \mathscr{J}^{\rm imp}_\ell(-p))^2 \leq ({\rm Im} \mathscr{J}^{\rm imp}_\ell(p))^2$. It then follows that the modulus of the improved $S$-matrix for self-scattering, defined as $s_\ell^{\rm imp}(p) = \mathscr{J}^{\rm imp}_\ell(-p)/\mathscr{J}^{\rm imp}_\ell(p)$, is bounded from above:
\begin{align}
    |s^{\rm imp}_\ell(p)| = \sqrt{\frac{1+({\rm Im}\mathscr{J}_\ell^{\rm imp}(-p))^2/({\rm Re} \mathscr{J}_\ell^{\rm imp}(p))^2}{1+({\rm Im}\mathscr{J}_\ell^{\rm imp}(p))^2/({\rm Re} \mathscr{J}_\ell^{\rm imp}(p))^2}} \leq 1.
\end{align}
This bound on the self-scattering $S$-matrix ensures that the improved annihilation cross section given in Eq.~(\ref{eq:improved SE}) satisfies the unitarity constraint Eq.~(\ref{eq:unitarity bound}), demonstrating that the RG improvement consistently restores unitarity even in the presence of shallow bound states.

By employing the low-energy expansion of the Jost function, we can investigate the detailed behavior of the improved amplitude $f_\ell^{\rm imp}(p)$ and the improved annihilation cross section $(\sigma_\ell v)^{\rm ann}$ in the limit $p \to 0$. As in the previous section, expanding the Jost function for the long-range potential as $\mathscr{J}_\ell(p) \simeq \alpha_\ell+\beta_\ell p^2 + i\gamma_\ell p^{2\ell+1}$, the improved Jost function becomes
\begin{align}
    \label{eq:improved jost expansion}
    \mathscr{J}^{\rm imp}_\ell(p) \simeq   \alpha_\ell + \beta_\ell p^2 + i\gamma_\ell p^{2\ell+1} + \frac{f_\ell^S(p)}{\gamma_\ell p^{2\ell}} .
\end{align}
Taking into account that the UV amplitude is proportional to $p^{2\ell}$, one can find that the constant term $\alpha_\ell$ in the Jost function receives a shift due to the improvement. In particular, the contribution from the imaginary part of $f_\ell^S(p)$, i.e., the  annihilation effects, renders this shift complex. It is evident that this shift regulates the improved Sommerfeld factor in the limit $p\to 0$, $S_\ell^{\rm imp,max}(p)=|\alpha_\ell+f_\ell^{S}(p)/\gamma_\ell p^{2\ell}|^{-2}$, and even in the zero-energy resonance limit $\alpha_\ell \to 0$, $S_\ell^{\rm imp}(p)$ approaches a constant value for all partial waves. 

This observation provides a simple prescription for the improved Sommerfeld enhancement in the small-momentum regime, especially, for the $s$-wave case, as originally proposed in\,\cite{Blum:2016nrz}. Neglecting the real part of $f^S(p)$, the improved Jost function can be approximated as $\mathscr{J}_\ell^{\rm imp}(p) \simeq \alpha_0 + i\gamma_0(p+p_c) \simeq \mathscr{J}_\ell(p+p_c)$,
with $p_c$ being a regularising momentum given by $p_c = {\rm Im} f_\ell^S(p)/\gamma_0^2 = \mu (\sigma_0v)^{\rm ann,0}/4\pi \gamma_0^2$. Therefore, the improved Sommerfeld factor can also be approximated by the naive one evaluated at a momentum shifted by the critical value, $S^{\rm imp}_0(p) \simeq S_0(p+p_c)$. In the Appendix~\ref{app: numerical}, we explicitly verify, that the improvement achieved by the RG method agrees well numerically with \cite{Blum:2016nrz} and its extension \cite{Parikh:2024mwa}.

The RG improvement applied to the scattering amplitude also modifies the effective range formula in Eq.~(\ref{eq:effective range expansion}). Using the expansion in Eq.~(\ref{eq:improved jost expansion}) we obtain the following expansion for the improved phase shift $\delta^{\rm imp}_\ell(p) = -i\ln s^{\rm imp}_\ell(p)/2$:
\begin{align}
     p^{2\ell+1}\cot \delta^{\rm imp}_{\ell}(p) = -\frac{1}{a_\ell} + \frac{r_\ell}{2}p^2 - \frac{f_\ell^S(p)}{\gamma_\ell^2p^{2\ell}} + \cdots.
\end{align}
The above expression implies that the scattering length $a_\ell$ is shifted by the influence of the UV amplitude through the improvement. In particular, due to the positivity of the imaginary part of the UV amplitude, ${\rm Im} f^S_\ell(p)>0$, the imaginary part of the shift in the inverse scattering length, $-1/a_\ell \to -1/a_\ell -f_\ell^S(p)/\gamma_\ell p^{2\ell}$, is positive. Therefore, the RG improvement reproduces the low-energy behavior of the scattering amplitude anticipated in Sec.~\ref{Cause of unitarity violation} within the zero-range approximation, and is consequently expected to induce instability of the bound states, as discussed next.

Since the RG improvement modifies the Jost function, its zeros—namely, the pole positions of the improved amplitude—are shifted. From a physical standpoint, this shift must be associated with the appearance of a decay width for the bound states due to annihilation. Let us examine the zeros of the improved Jost function in Eq.~(\ref{eq:improved jost}) to verify this connection. Denote the position of the zero of the original Jost function $\mathscr{J}_\ell(p)$ as $p=p_B=i\kappa \,(\kappa>0)$, corresponding to the bound state with its binding energy $E_B = -\kappa^2/2\mu$, and that of the improved Jost function $\mathscr{J}^{\rm imp}_\ell(p)$ as $p=p_B-\gamma$. Then the shift $\gamma$, to its first order, satisfies the following relation:
\begin{align}
    \label{eq:BS 1}
     - \gamma \left. \frac{d}{dp}\mathscr{J}_\ell(p)\right|_{p=p_B} + \frac{\mu p_B}{2\pi(2\ell+1)}\frac{h_\ell p_B^{2\ell}}{\mathscr{J}_\ell(-p_B)} = 0.
\end{align}
Here we considered only the imaginary part of the UV amplitude $f^S_\ell(p) \simeq i\mu (\sigma_\ell v)^{\rm ann,0} / 4\pi(2\ell+1)$ and parametrized the annihilation cross section as $ (\sigma_\ell v)^{\rm ann,0} = h_\ell p^{2\ell}$. It is known that the derivative of the Jost function at the zero corresponding to the bound state satisfies the following relation (see Appendix.~\ref{app: Jost} for details):
\begin{align}
    \label{eq:BS 2}
    \left. \frac{d}{dp}\mathscr{J}_\ell(p)\right|_{p=p_B} = \frac{i(-1)^\ell\mathscr{J}_\ell(-p_B)}{\mathcal{A}_\ell^2},
\end{align}
where $\mathcal{A}_\ell$ is the so-called asymptotic normalization constant, which corresponds to the normalization of the reduced radial wave function $u_{\ell,p_B}(r)$ (i.e., the radial wave function $R_{\ell,p_B}(r)$ multiplied by $r$, $R_{\ell,p_B}(r) = u_{\ell,p_B}(r)/r$) of the bound state at asymptotically large distances, $u_{\ell,p_B}(r) \to \mathcal{A}_\ell e^{-\kappa r}$. The reduced wave function is taken to be real and normalized to be 1 as $\int^\infty_0 dr\, (u_{\ell,p_B}(r))^2=1$. By comparing the asymptotic behavior of the reduced wave function $u_{\ell,p_B}(r)$ with that of the regular solution at large distances in Eq.~(\ref{eq:reg asymptotic}), $\phi_{\ell,p_B}(r) \to -i\mathscr{J}_\ell(-p_B)h^+_\ell(p_Br)/2$, the asymptotic normalization constant $\mathcal{A}_\ell$ can further be related to the higher-order derivatives of the radial wave function $R_{\ell,p_B}(r)$ at the origin as
\begin{align}
    \label{eq:BS 3}
    \left.\frac{d^\ell}{dr^\ell}R_{\ell,p_B}(r)\right|_{r=0} = \frac{2i^{\ell+1} \mathcal{A}_\ell}{\mathscr{J}_\ell(-p_B)} \frac{\ell!}{(2\ell+1)!!}p_B^{\ell+1}.
\end{align}
By combining Eqs.~(\ref{eq:BS 1}), (\ref{eq:BS 2}), and (\ref{eq:BS 3}), we find that the bound state corresponding to the zero of the Jost function at $p=p_B$ acquires imaginary contribution to its binding energy, $E_B' = (p_B-\gamma)^2/2\mu \simeq E_B -i\kappa \gamma/\mu $, which in turn leads to the following decay width:
\begin{align}
    \Gamma = \frac{h_\ell}{4\pi(2\ell+1)}\left[\frac{(2\ell+1)!!}{\ell!}\left.\frac{d^\ell}{dr^\ell}R_{\ell,p_B}(r)\right|_{r=0}\right]^2.
\end{align}
The above expression correctly reproduces the expected formula for the decay width associated with the annihilation of a bound state\,\cite{Petraki:2015hla,Petraki:2025zvv}. In particular, the decay width of the $s$-wave bound state takes the well known form $\Gamma = (\sigma_0 v)^{\rm ann,0} |\Psi_0(0)|^2$ with $\Psi_0(r)$ being the total wave function of the bound state, given by the product of radial wave function and the $s$-wave spherical harmonics. This demonstrates that the RG-improved amplitude Eq.~(\ref{eq:RG improved amplitude simplified}) correctly captures the bound-state poles, including their decay widths, while simultaneously achieving the unitarization of the Sommerfeld effect.

Let us illustrate the effectiveness of our proposed unitarization method through a simple example. We consider the spherical well potential
\begin{align}
\label{eq:well}
V(r) = -\frac{p_V^2}{2\mu}\theta(R - r),
\end{align}
where $p_V>0$ parametetrizes the depth of the well and $R$ denotes its width. To determine the Jost function, one solves the Schr\"{o}dinger equation with the potential given in Eq.~(\ref{eq:well}) under the boundary condition that the solution coincides with the Riccati–Bessel function $j_\ell(pr)$ as $r\to0$, examines its asymptotic behavior at infinity, and compare it with Eq.~(\ref{eq:reg asymptotic}). For the well potential, all these procedures can be carried out analytically, leading to the result
\begin{align}
    \label{eq:well-jost}
    \mathscr{J}_\ell(p) = \left(\frac{p}{\tilde{p}}\right)^\ell\left[j'_\ell(\tilde{p}R)h^+_\ell(pR)-\frac{p}{\tilde{p}}j_\ell(\tilde{p}R){h^+_\ell}'(pR)\right].
\end{align}
Here, $\tilde{p}$ is given by $\tilde{p}=\sqrt{p^2+p_V^2}$, and $j_\ell'(x)$ and $h_\ell'(x)$ denote the derivatives of Riccti-Bessel and Riccati-Hankel functions, respectively. Using the Jost function in Eq.~(\ref{eq:well-jost}) together with Eqs.~(\ref{eq:analytic s matrix}) and (\ref{eq:def SE}), the long-range self-scattering amplitude $f^L_\ell(p)$ and the conventional Sommerfeld factor $S_\ell(p)$ are obtained. When the well depth, i.e.,  $p_V$, takes specific values, the Jost function has a zero at $p = 0$, indicating the emergence of a zero-energy resonance in the system. For the $s$-wave ($\ell = 0$) and the $p$-wave ($\ell = 1$), these conditions are given by
\begin{align}
    p_V R =
    \left\{
        \begin{aligned}
            &\pi/2, 3\pi/2, 5\pi/2, \dots \,(\ell=0)\\
            &\pi, 2\pi, 3\pi, \dots \,(\ell=1)
        \end{aligned}
    \right.
    .
\end{align}
As $p_V$ approaches these critical values, the Sommerfeld factor $S_\ell(p)$ becomes significantly enhanced as $p \to 0$, exhibiting incompatible growth with the unitarity bound, and it requires unitarization. In addition to the Jost function, the UV amplitude $f^S_\ell(p)$ is also required for the RG-improved scattering amplitude and Sommerfeld factor in Eqs.~(\ref{eq:improved jost}) and (\ref{eq:improved SE}). As in the previous sections, we adopt the approximation in which only its imaginary part is retained, $f^S_\ell(p) \simeq i\mu (\sigma_\ell v)^{\rm ann,0} / 4\pi(2\ell+1)$, and assume that the tree-level annihilation cross section for the $\ell$-th partial wave $(\sigma_\ell v)^{\rm ann,0}$ takes the following form:
\begin{align}
    \label{eq:tree level annihilation}
    (\sigma_\ell v)^{\rm ann,0} = \frac{4\pi(2\ell+1) \alpha_D^2}{\mu^2}\left(\frac{p}{\mu}\right)^{2\ell},
\end{align}
where $\alpha_D$ is the coupling constant characterizing the strength of annihilation.

\begin{figure}[t]
    \centering
    \includegraphics[width=0.49\linewidth]{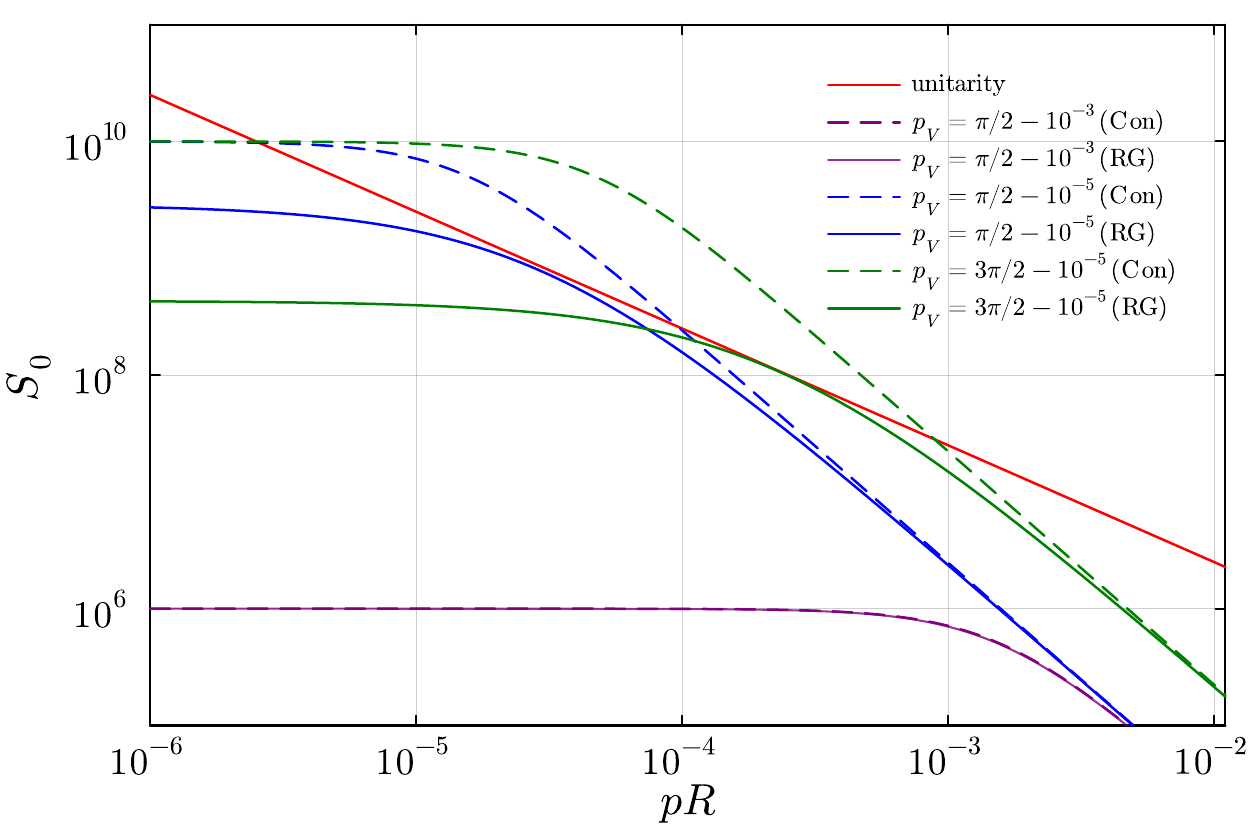}
    \includegraphics[width=0.49\linewidth]{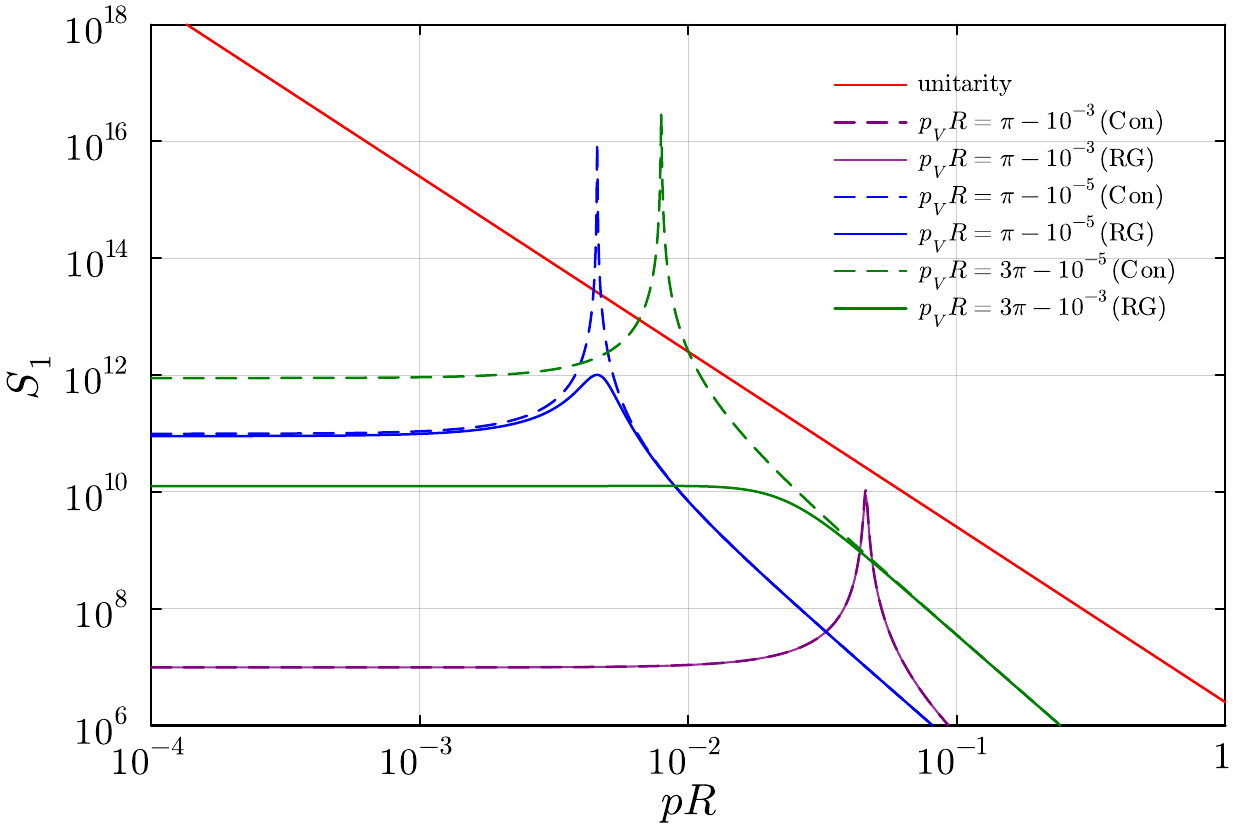}
    \caption{\small \sl 
    Sommerfeld factors as functions of momentum for the $s$-wave (left panel) and $p$-wave (right panel) with the spherical well potential. The dashed lines correspond to the conventional calculation (Con), while the solid lines represent the RG-improved results (RG). The red line indicates the upper bound imposed by unitarity. The mass parameter and the coupling constant are take to be $\mu R = 10$ and $\alpha_D = 10^{-2}$, respectively. See the main text for details.
    }
    \label{fig: well_SE}
\end{figure}

\begin{figure}[t]
    \centering
    \includegraphics[width=0.495\linewidth]{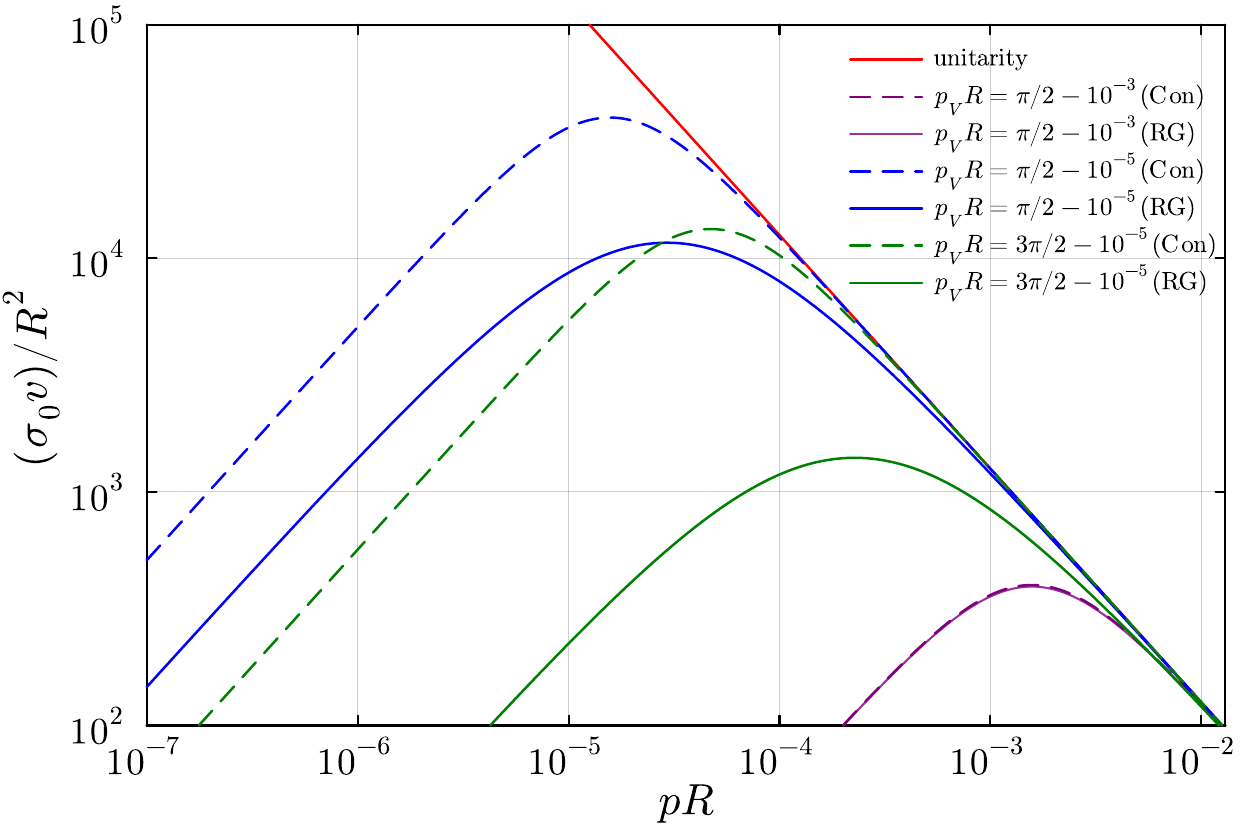}
    \includegraphics[width=0.495\linewidth]{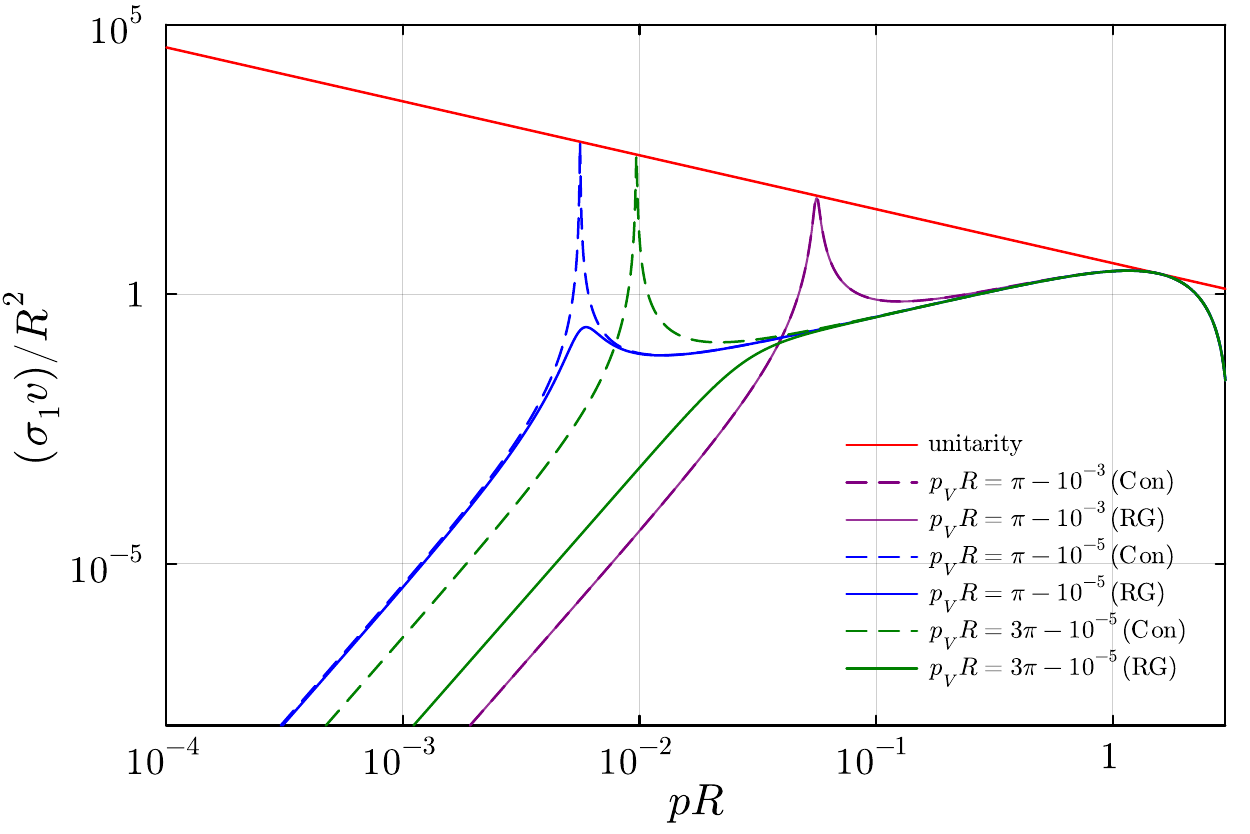}
    \caption{\small \sl 
    Self-scattering cross sections as functions of momentum for the $s$-wave (left panel) and $p$-wave (right panel) with the spherical well potential. The dashed lines correspond to the conventional calculation (Con), while the solid lines represent the RG-improved results (RG). The red line shows the unitarity bound for the self-scattering cross section. The mass parameter and the coupling constant are take to be $\mu R = 10$ and $\alpha_D = 10^{-2}$, respectively. See the main text for details.
    }
    \label{fig: well_el}
\end{figure}

Fig.~\ref{fig: well_SE} shows the Sommerfeld factors as functions of momentum $p$ (normalized by the well width $R$) for the $s$-wave (left panel) and the $p$-wave (right panel). We set the mass parameter and the coupling constant to $\mu R = 10$ and $\alpha_D = 10^{-2}$, respectively. In both panels, the dashed and solid lines of the same color form a pair: the dashed line represents the conventional Sommerfeld factor $S_\ell(p)$ given in Eq.~(\ref{eq:def SE}), while the corresponding solid line indicates the RG-improved Sommerfeld factor $S_\ell^{\rm imp}(p)$ in Eq.~(\ref{eq:improved SE}). The red solid line shows the ratio of the unitarity-bound annihilation cross section, which is given by the RHS of Eq.~(\ref{eq:unitarity bound}), to the tree-level annihilation cross section Eq.~(\ref{eq:tree level annihilation}). For the $s$-wave case, we plot the Sommerfeld factors with the well depth chosen near the first zero-energy resonance, at $p_V = \pi/2 - 10^{-3}$ (purple), $p_V = \pi/2 - 10^{-5}$ (blue), and near the second zero-energy resonance, at $p_V = 3\pi/2 - 10^{-5}$ (green). For the purple line, which is sufficiently far from the zero-energy resonance, the conventional Sommerfeld factor remains well below the unitarity bound and agrees well with the improved one. In contrast, for the blue and green lines, which are closer to the zero-energy resonance, both show unphysical behavior at small momenta, exceeding the unitarity bound. The RG-improved Sommerfeld factor resolves these issues and yields results consistent with unitarity. For the $p$-wave case, we plot the points $p_V = \pi - 10^{-3}$ (purple) and $p_V = \pi - 10^{-5}$ (blue) near the first zero-energy resonance, and $p_V = 3\pi - 10^{-5}$ (green) near the third zero-energy resonance. Unlike the $s$-wave case, the conventional Sommerfeld factors for all three parameter choices exhibit spike-like structures\,\cite{Beneke:2024iev}. These arise from the fact that the $p$-wave can support resonant states, corresponding to zeros of the Jost function in the region of positive momentum. In particular, the blue and green lines exhibit resonant enhancements that exceed the unitarity bound. In contrast, the RG-improved Sommerfeld factor, shown as dashed lines, resolves such unphysical behavior: while it coincides with the conventional one for the purple line, which does not violate unitarity, the resonance peaks in the blue and green lines are properly regulated, yielding results consistent with unitarity.

The RG improvement modifies not only the Sommerfeld factor but also the self-scattering cross section. Fig.~\ref{fig: well_el} shows the self-scattering cross section as a function of momentum for the $s$-wave (left panel) and $p$-wave (right panel).  We again set the mass parameter and the coupling constant to $\mu R = 10$ and $\alpha_D = 10^{-2}$, respectively. The dashed lines represent the conventional self-scattering cross sections, while the solid lines correspond to the RG-improved results. The color scheme follows that of Fig.~\ref{fig: well_SE}: for the $s$-wave, purple corresponds to $p_V = \pi/2 - 10^{-3}$, blue to $p_V = \pi/2 - 10^{-5}$, and green to $p_V = 2\pi - 10^{-5}$; for the $p$-wave, purple corresponds to $p_V = \pi - 10^{-3}$, blue to $p_V = \pi - 10^{-5}$, and green to $p_V = 3\pi - 10^{-5}$. In addition, the red solid line indicates the unitarity bound for the self-scattering cross section, which is four times the corresponding bound for annihilation. For both the $s$- and $p$-wave cases, as in Fig.~\ref{fig: well_SE}, the purple line, which corresponds to a parameter point far from the zero-energy resonance, shows good agreement between the conventional calculation and the RG-improved result. In contrast, for the blue and green lines, which are closer to the zero-energy resonance, the conventional results remain consistent with the unitarity bound in both cases, but the RG improvement suppresses the peaks. In particular, the resonance structures that appear in the $p$-wave case are smeared out due to the inclusion of annihilation effects.

\section{Discussion}
\label{sec: Discussion}

In this section, we provide further insights into the results obtained in the previous section by addressing two complementary aspects. First, we revisit the RG improvement procedure from the perspective of Wilsonian renormalization group\,\cite{Wilson:1973jj}, clarifying the validity and limitations of our method. This analysis helps identify conditions under which the RG improvement may fail or require modification. Second, as an advanced topic concerning the Sommerfeld effect, we briefly discuss the bound-state formation process\,\cite{An:2016gad,vonHarling:2014kha,Asadi:2016ybp}, which can exhibit behaviors that violate the unitarity bound and therefore require careful treatment. These discussions aim to deepen the understanding of our formalism and its potential applicability, though they remain at a qualitative level and a more detailed treatment is left for future research.

\subsection{Insights from Wilsonian renormalization group}
\label{sec: Wilson}

In Section~\ref{sec:Sommerfeld enhancement}, we treated the short range potential $V_S$ as a perturbation and derived the conventional Sommerfeld factor. In the following Section~\ref{sec:RG Sommerfeld}, we obtained the improved Sommerfeld factor by resumming the secular terms that appear in the perturbative expansion of the self-scattering amplitude. These results can be characterized within the framework of the exact renormalization group approach for distorted waves, as introduced in\,\cite{Barford:2002je}.

By constructing the Wilsonian renormalization group for the distorted-wave Lippmann-Schwinger equation, one can classify short-range potential $V_S$. Specifically, by introducing a momentum cutoff $\Lambda$ as the upper limit of the momentum integral in the momentum-space representation of the equation, and requiring that the full amplitude be independent of $\Lambda$, one obtains a renormalization group equation that governs the $\Lambda$-dependence of $V_S(\Lambda)$:
\begin{align}
    \label{eq:RG short range potential}
    \frac{\partial V_S}{\partial \Lambda} = -V_S \frac{\partial G_L}{\partial \Lambda} V_S,
\end{align}
where $G_L$ is a Green function associated with the long-range potential $V_L$. (For its explicit form, see\,\cite{Barford:2002je}; however, the details are not essential for the following discussion.) The quantity $V_S(\Lambda)$ can be interpreted as the effective potential at the scale $p \sim \Lambda$, and the renormalization group equation describes its flow as a function of the momentum scale. The fixed points of the renormalization group equation (\ref{eq:RG short range potential}) and their stability in the limit $p \to 0$ characterize the nature of the short-range contributions to the scattering process. 

It is evident that the equation admits a trivial fixed point $V_S = 0$, which describes a system with only distorted waves of the long-range potential $V_L$. Moreover, a detailed analysis reveals that this fixed point is stable: all perturbations around it are irrelevant, and they flow into the fixed point as $p \to 0$. This observation provides a justification for our previous treatment of the short-range potential $V_S$ throughout this paper. In our evaluation of the Sommerfeld factor, $V_S$ was determined through perturbative matching with some underlying UV theory, under the assumption that its contribution to the scattering amplitude is sufficiently small. This implies that $V_S$ serves as a small perturbation to the scattering primarily governed by the long-range potential $V_L$. From the viewpoint of the exact renormalization group, this situation corresponds precisely to a perturbation around the trivial fixed point—namely, the fixed point representing pure long-range scattering without short-range modifications. Irrelevant interactions do not grow under the flow toward low energies, which justifies their perturbative treatment. The only challenge arises from the secular terms associated with the unperturbed part, but these can be removed through renormalization. This discussion is reminiscent of perturbative field theories such as $\phi^4$ theory and QED in four dimensions. These theories are formulated around the Gaussian fixed point, which is a trivial fixed point, and all their interactions are irrelevant. Consequently, at low energies, it is justified to describe physical observables using perturbative expansions.

On the other hand, the renormalization group equation also allows for the existence of nontrivial fixed points. In cases such as the Yukawa interaction, considered here as an example of a long-range force, the corresponding fixed point is unstable. That is, there exist relevant directions in the space of short-range interactions that grow as $p \to 0$. As a result, if $V_S$ has components along these directions, the perturbative expansion in $V_S$ becomes invalid in the low-energy regime. Situations in which the underlying UV theory gives a sizable contribution to the self-scattering process at low energy fall into this class of problems. In such cases, the short-range amplitude can no longer be treated as a perturbation, and the perturbative expansion around the long-range interaction breaks down. As briefly mentioned in Sec.~\ref{sec:RG Sommerfeld}, in such cases renormalized perturbation theory also ceases to be effective. In analogy with field theory, these scenarios are similar to QCD, where the coupling is a relevant interaction: perturbative QCD offers a reliable description at high energies, but fails in the low-energy regime.

In situations where the contribution to the scattering from the underlying UV theory is significant and the perturbative expansion with respect to $V_S$ is no longer applicable, the modified effective range expansion\,\cite{Barford:2002je,vanHaeringen:1981pb}, which generalizes the effective range expansion in Eq.~(\ref{eq:effective range expansion}) to systems with both long-range and short-range interactions, provides a good fitting formula:
\begin{align}
    \label{eq:modified ERE}
    \frac{1}{|\mathscr{J_\ell}(p)|^2}p^{2\ell+1}[\cot \tilde{\delta}_\ell^S(p)-i]+\mathcal{M}_\ell(p) = -\frac{1}{\tilde{a}_\ell} + \frac{\tilde{r}_\ell}{2}p^2 + \cdots,
\end{align}
where $\tilde{\delta}^S_\ell(p)$ is defined by the difference between the total phase shift $\delta_\ell(p)$ and that due to the long-range potential $\delta^L_\ell(p)$. $\mathcal{M}_\ell(p)$ is given by $\mathcal{M}_\ell(p) = \lim_{r \to 0}(d/dr)^{2\ell+1}((pr)^\ell \chi^+_{\ell,p}(r))/2^\ell \ell! \mathscr{J}_\ell(p)$, which is constructed using the Jost solution $\chi^+_{\ell,p}(r)$ (the solution of the Schr\"{o}dinger equation (\ref{eq:radial sceq}) with boundary conditions imposed at infinity, $\chi^+_{\ell,p}(r)/h^+_\ell(pr) \to 1, r \to \infty$). The extra factors appearing in the modified effective range expansion Eq.~(\ref{eq:modified ERE}) compared to the standard one Eq.~(\ref{eq:effective range expansion}) serve to remove the non-analytic momentum dependence arising from the long-range potential. This expansion has been used to subtract long-range effects such as the Coulomb force or one-pion exchange between nucleons\,\cite{Steele:1998zc,Bethe:1949yr}, thereby enabling the extraction of low-energy properties of the strong interaction. Therefore, if one allows the shape parameters $\tilde{a}_\ell, \tilde{r}_\ell$ to take complex values to account for annihilation processes, Eq.~(\ref{eq:modified ERE}) is expected to provide a useful parametrization of the self-scattering amplitude in situations where the short-range effect is nonperturbative, though it remains applicable even when the short-range effect is perturbative.

The unitarization of the Sommerfeld enhancement essentially requires the resummation of short-range effects. In our approach, we carried out this resummation using the RG method. However, there exists an alternative approach, which involves solving the Schr\"{o}dinger equation including the short-range potential $V_S$\,\cite{Blum:2016nrz,Flores:2024sfy,Parikh:2024mwa}. In fact, the derivation of the conventional Sommerfeld enhancement proceeds by solving the Schrödinger equation with a long-range potential $V_L$ determined from the UV theory via the Born approximation, thereby resumming the long-range effects. However, this method must be applied with considerable care. This is because the short-range potential is generally represented by distributions such as delta functions or their derivatives, and the wavefunctions corresponding to such potentials are also distributions. Since the Schr\"{o}dinger equation involves the product of these distributions, regularization becomes unavoidable in order to give meaning to these products.

Once the potential is regularized by some method and the Schr\"{o}dinger equation is solved, the resulting $S$-matrix must reduce to the form of the modified effective range expansion, as it provides a general parametrization of the scattering amplitude for low-energy regime in the case where the potential comprises two separate contributions. In this case, the shape parameters become functions of the regularization parameters (and parameters contained in the long-range potential as well), but they are fixed to definite numerical values by matching to physical observables—namely, through renormalization. This approach is, in essence, equivalent to using the modified effective range expansion and fitting its parameters to scattering data at a suitable momentum scale. The key point, then, is whether the effect of the short-range potential is perturbative or nonperturbative. If the short-range interaction is perturbative, the corresponding contribution to the scattering amplitude can be computed directly from the UV theory using perturbation theory. In such cases, the RG–based treatment we have proposed is expected to provide a sufficiently accurate and much simpler approximation, particularly for computing unitarized Sommerfeld factors,  compared to performing a matching based on the modified effective range expansion. Indeed, Ref.~\cite{Blum:2016nrz} discusses that, for the $s$-wave case, solving the Schr\"{o}dinger equation with the short-range effect included yields results that agree well numerically with our expression given in Sec.~\ref{sec:RG Sommerfeld} when the underlying short-range effect is perturbative, and we also confirm in Appendix~\ref{app: numerical} that the RG method and the direct solution of the Schr\"{o}dinger equation in \cite{Blum:2016nrz,Parikh:2024mwa} agree well numerically for the spherical well potential. 

On the other hand, the analytical connection between the RG method and the formulations of Refs.~\cite{Blum:2016nrz} and \cite{Parikh:2024mwa} is conceptually subtle. Those works define short-range interactions—represented by delta functions and their derivatives—nonperturbatively and analyze the Sommerfeld effect, which leads them to encounter UV divergences immediately. By contrast, our framework defines the short-range interaction only at a perturbative level as in Eq.~(\ref{eq:DWBA simplified}); as a result, the UV divergences appear sequentially with increasing perturbative order. Since in this work we restrict attention to the lowest-order approximation with UV-finite short-range input, we do not face those UV divergences. While extending our treatment to higher perturbative orders would indeed reintroduce UV issues, our view is that the essential unitarity question for the Sommerfeld effect is infrared (IR) rather than UV in nature; within that perspective, our method addresses unitarity correctly at the level considered. Establishing a more general analytical correspondence between the approaches based on exact solutions of the Schr\"{o}dinger equation or on the modified effective range expansion and the RG-based framework—particularly in cases where the short-range interaction is perturbative for arbitrary partial waves—remains an important direction for future work.

\subsection{Bound state formation}

In the preceding discussion, we have considered only the simple process of direct annihilation. As a higher-order contribution to the annihilation process, it is known that slowly moving particles can form bound states through the emission of soft bosons that mediate their long-range interactions, which is referred to as bound state formation (BF)\,\cite{An:2016gad,vonHarling:2014kha,Asadi:2016ybp}. It has been pointed out that BF also exhibits resonance peaks at parameter values where the self-scattering cross section and the Sommerfeld factor resonate, in contradiction with the partial-wave unitarity bound\,\cite{Oncala:2019yvj,Binder:2023ckj}. The unitarity violation problem in BF with unequal potentials in the initial and final states is discussed more generally in\,\cite{Beneke:2024nxh}.

It would be interesting to apply the RG method to BF. Let us consider the simplest BF, where the mediator particle is a scalar boson $\varphi$, and that the transition from the scattering state $\mathcal{S}$ to a bound state $\mathcal{B}$ proceeds via the emission of a light boson with momentum $k$, mediated by a scalar transition operator $V_k$. At low energies, the initial state is dominated by the $s$-wave contribution and then, under the leading-order DWBA, the matrix element for this process is roughly given by
\begin{align}
    \label{eq: BF approximation}
    \mathcal{M}(\mathcal{S} 
    \to \mathcal{B} + \varphi) \simeq \braket{\mathcal{B}|V_k|\mathcal{S}}
    \propto
    \frac{1}{\mathscr{J}_0(0)}\int^\infty_0 dr\,\psi^{\mathcal{B}}(r)V_{k}(r)\phi_{0,p}(r)
\end{align}
where $\psi^{\mathcal{B}}(r)$ and $\phi_{0,p}(r)/\mathscr{J}_0(p)$ are the radial wave function for the bound state and the scattering state, respectively. This simple estimate clearly demonstrates that the amplitude can be substantially amplified by the inverse of the Jost function, especially in the presence of shallow bound states and narrow resonances. As noted in the previous chapter, this suggests that the lowest-order DWBA evaluation of BF may violate unitarity, calling for a resummation to restore consistency.

\begin{figure}[t]
    \centering
    \includegraphics[width=0.60\linewidth]{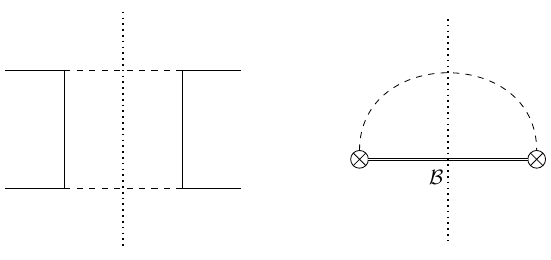}
    \caption{\small \sl 
    Diagrams for scattering processes contributing to self-scattering, induced from the UV theory. The left diagram shows the contribution from a direct decay process, while the right diagram depicts a contribution from a vertex mediating a transition to a bound state via the emission of a light boson. Solid lines denote the scattering particles, dashed lines their daughter particles, and double lines the bound states formed by the scattering particles. The dotted vertical lines represent Cutkosky cuts, indicating that these diagrams have nonzero imaginary parts.
    }
    \label{fig: BF}
\end{figure}

If we consider the above calculation of the BF process within the framework of the Sommerfeld factor discussed in Sec.~\ref{sec:Sommerfeld enhancement}, we may expect that the unitarization proceeds in the following way. Namely, the presence of the interaction that allows the scattering state to transition into a bound state $\mathcal{B}$ via the emission of a boson $\varphi$ introduces a new contribution to the UV amplitude $f^S_{\ell}(p)$ of the self-scattering process, which has been treated perturbatively. This process is illustrated in Fig.~\ref{fig: BF}. In this figure, the left diagram represents the self-scattering mediated by direct emission of $\varphi$s through annihilation, while the right diagram shows the additional contribution arising from the newly introduced BF vertex (depicted by a cross dot), where the intermediate state includes the bound state $\mathcal{B}$ and the emitted boson $\varphi$. The potential singularity of self-scattering amplitude arising from the consideration of this diagram originates from the factor of the Jost function associated with the external legs. Therefore, it constitutes a secular term, and by incorporating the BF effect into the UV amplitude $f_\ell^S(p)$, the annihilation cross section can be unitarized by means of Eq.~(\ref{eq:improved jost}).

It goes without saying that the above treatment of the BF process is highly naive and requires further work. For instance, if the aim is merely to suppress the unphysical enhancement from shallow bound states or narrow resonances, it may suffice—at least approximately—to replace the Jost function in Eq.~(\ref{eq: BF approximation}) with the improved version in Eq.~(\ref{eq:improved jost}), which incorporates only the tree-level annihilation effect. If the emitted boson is spin-1 and involves transitions to states with different angular momenta, a more refined approach based on coupled-channel analysis may be needed. There, the Jost function would be replaced by its multi-channel counterpart, the Jost matrix, and one would examine whether our unitarization procedure still applies. It is also possible that scattering is dominated by an intermediate state $\mathcal{B} + \varphi$, driving the theory to a nontrivial fixed point and rendering renormalized perturbation theory around the Gaussian fixed point ineffective. In that case, one might instead solve the Schrödinger equation directly or use the expansion in Eq.~(\ref{eq:modified ERE}). Under any of the above scenarios, these considerations are highly model-dependent, making it difficult to develop a general argument.

\section{Conclusion}
\label{sec: Conclusion}

In this work, we proposed a method to improve the Sommerfeld enhancement in a single-channel system under the assumption that the annihilation process can be treated perturbatively, in a way that is consistent with unitarity. To clarify the source of unitarity violation, we formulated the Sommerfeld enhancement using effective quantum mechanics along with certain advanced tools from scattering theory. This analysis revealed that the conventional perturbative expansion used to compute the Sommerfeld enhancement becomes a singular perturbation when the spectrum induced by the long-range interaction contains shallow bound states or narrow resonances. This leads to the appearance of secular terms. Therefore, in order to ensure unitarity, a resummation of higher-order contributions is required.

We carried out this resummation using the renormalization group approach. This method requires, in addition to the conventional Sommerfeld factor, input from the UV scattering amplitude and the phase shift of the self-scattering induced by the long-range interaction. The resulting scattering amplitude respects unitarity and exhibits saturation of the excessive Sommerfeld enhancement at low energies caused by resonances. Furthermore, by analyzing the poles of the improved amplitude, we find that bound states acquire a finite decay width. This provides a concrete realization of a unitarization mechanism that was anticipated in the original work on Sommerfeld enhancement.

In the discussion, we examined our approach from the viewpoint of the Wilsonian renormalization group. From this perspective, our method under the assumption of a perturbative UV theory can be understood as a perturbative expansion around the Gaussian fixed point. The improvement of the amplitude through perturbative renormalization shares a parallel structure with the standard renormalized perturbation theory applied to weakly coupled field theories such as QED. Hence, the renormalization group approach is expected to offer a more tractable framework for unitarizing the Sommerfeld enhancement in systems with weak short-range interactions, compared to approaches based on exact solutions to the Schr\"{o}dinger equation. Although the RG method shows good numerical agreement with that approach, establishing a precise analytical correspondence between the two frameworks for arbitrary partial waves remains an open problem.

We also discussed bound-state formation, which has been widely investigated as a higher-order annihilation effect. The standard calculation of bound-state formation is also a singular perturbation problem and thus requires resummation. We expect that our method remains effective in this case as well, provided that the bound-state formation is properly incorporated as an additional contribution to the UV process.


\section*{Acknowledgements}

The author would like to express sincere gratitude to Shigeki Matsumoto for carefully reading the draft and providing valuable comments.

\appendix

\section{Bessel functions}
\label{app: Bessel}

In this appendix, we summarize the properties of the types of Bessel functions that appear in partial wave analysis in scattering theory. Section~\ref{app:Riccati-Bessel} discusses the Riccati–Bessel functions, which are solutions to the free radial Schr\"{o}dinger equation with zero potential $V(r)=0$. Section ~\ref{app:Coulomb-Bessel} describes the solutions to the radial Schr\"{o}dinger equation and the corresponding Jost function in the presence of a Coulomb potential $V(r) = -\alpha/r$ with $\alpha$ being the fine-structure constant, which was not treated in the main text. The primary purpose here is to fix the notation of these special functions as used in this paper. Therefore, we do not provide detailed derivations, and the reader is referred to standard textbooks for such materials.

\subsection{Riccati-Bessel functions}
\label{app:Riccati-Bessel}

We describe the solutions of the Schr\"{o}dinger equation without potential $V(r)=0$. The radial free Schrödinger equation
\begin{align}
    \label{App:Bessel-schro}
    \left[\frac{d^2}{dr^2}-\frac{\ell(\ell+1)}{r^2}+p^2\right]u_{\ell,p}(r) = 0
\end{align}
is a second-order ordinary differential equation with a regular singular point at $r=0$. Accordingly, one of the two linearly independent solutions can be chosen to be regular at the origin. Introducing the variable $z=pr$, we select, among the two linearly independent solutions, the Riccati–Bessel $j_\ell(z)$ and Riccati–Neumann functions $n_\ell(z)$ as the regular and irregular solutions, respectively, defined by
\begin{align}
    j_\ell(z) = z^{\ell+1}\sum_{n=0}^\infty \frac{(-1)^n}{n!(2\ell+2n+1)!!}\left(\frac{z^2}{2}\right)^n, \quad n_\ell(z) = \frac{1}{z^\ell}\sum^\infty_{n=0}\frac{(-1)^n(2\ell-2n-1)!!}{n!}\left(\frac{z^2}{2}\right)^n.
\end{align}
The functions $j_\ell(z)$ and $n_\ell(z)$ behave near the origin as
\begin{align}
    \label{App:Bessel at the origin}
    j_\ell(z) \underset{z \to 0}{\longrightarrow} \frac{z^{\ell+1}}{(2\ell+1)!!}, \quad  n_\ell(z) \underset{z \to 0}{\longrightarrow} \frac{(2\ell-1)!!}{z^\ell},
\end{align}
and have the following asymptotic forms at infinity:
\begin{align}
    j_\ell(z) \underset{z \to \infty}{\longrightarrow} \sin(z-\ell \pi/2), \quad n_\ell(z)\underset{z \to \infty}{\longrightarrow}\cos(z-\ell \pi/2).
\end{align}
The following Riccati–Hankel functions, which are linear combinations of $j_\ell(z)$ and $n_\ell(z)$, are also solutions of Eq.~(\ref{App:Bessel-schro}) and useful:
\begin{align}
    h_\ell^\pm(z) = n_\ell(z) \pm i j_\ell(z).
\end{align}
Their asymptotic behavior at large distances is given by
\begin{align}
    h^\pm_\ell(z) \underset{z \to \infty}{\longrightarrow} e^{\pm i(z-\ell \pi/2)}.
\end{align}
The explicit forms of these Bessel functions for $\ell=0$ and $\ell=1$ are given as follows:
\begin{align}
    j_0(z) = \sin z, &\quad j_1(z) = \frac{\sin z}{z}-\cos z,\\
    n_0(z) = \cos z, &\quad n_1(z) =\frac{\cos z}{z} + \sin z,\\
    h^\pm_0(z) = e^{\pm i z}, &\quad h_1^{\pm}(z) =\left(1+\frac{i}{z}\right) e^{\pm i(z-\pi/2)}.
\end{align}

\subsection{Coulomb functions}
\label{app:Coulomb-Bessel}

We consider the following radial Schr\"{o}dinger equation under the Coulomb potential $V(r) = -\alpha/r$, where $\alpha$ is the fine-structure constant. The potential is attractive for $\alpha >0$ and repulsive for  $\alpha<0$:
\begin{align}
    \label{App: Coulomb schro}
    \left[
        \frac{d^2}{dr^2}-\frac{\ell(\ell+1)}{r^2}+\frac{2\eta p}{r} +p^2
    \right]
    u_{p,\ell}(r) = 0.
\end{align}
Here, $\eta$ denotes the Sommerfeld parameter, given by $\eta = \mu \alpha/p$ with $\mu$ being the mass. Eq.~(\ref{App: Coulomb schro}) is a second-order differential equation with a regular singular point at $r=0$, and can be reduced, via an appropriate change of variables, to the confluent hypergeometric differential equation. Among the two linearly independent solutions, we denote by $F_\ell(pr)$ the one that is regular at the origin, and by $G_\ell(pr)$ the one that is irregular. These are chosen so that they exhibit the following asymptotic behavior at infinity:
\begin{align}
    \label{App: Coulomb waves at infinity}
    F_\ell(pr) &\underset{r \to \infty}{\longrightarrow} \sin(pr+\eta \ln (2pr)-\ell \pi/2 + \sigma_\ell), \\
    G_\ell(pr) &\underset{r \to \infty}{\longrightarrow} \cos(pr +\eta \ln (2pr)-\ell \pi/2 + \sigma_\ell),
\end{align}
where $\sigma_\ell$ is the Coulomb phase shift, which is given by
\begin{align}
    \sigma_\ell = \frac{1}{2i}(\ln\Gamma(\ell+1-i\eta)-\ln \Gamma(\ell+1+i\eta)),
\end{align}
and corresponds to the phase shift associated with Coulomb scattering. In particular, for physical scattering with $p>0$, $\sigma_\ell = \arg \Gamma(\ell+1-i\eta)$. From Eq.~(\ref{App: Coulomb waves at infinity}), it is evident that the wave function under the Coulomb potential includes not only a phase shift $\sigma_\ell$ due to scattering but also a logarithmic distortion. This arises from the fact that the decay of the Coulomb potential at large distances is slower than that of the centrifugal potential.

The Coulomb wave functions $F_\ell(pr)$ and $G_\ell(pr)$ exhibit the following behavior near the origin:
\begin{align}
    \label{App: Coulomb waves at the origin}
    F_\ell(pr) \underset{r \to 0}{\longrightarrow} \frac{C_\ell(\eta)}{(2\ell+1)!!}(pr)^{\ell+1}, \quad 
    G_\ell(pr) \underset{r \to 0}{\longrightarrow} \frac{(2\ell-1)!!}{C_\ell(\eta)}(pr)^{-\ell}.
\end{align}
By comparing Eqs.~(\ref{App:Bessel at the origin}) and (\ref{App: Coulomb waves at the origin}), it is seen that $F_\ell(pr)$ and $G_\ell(pr)$ correspond to the Riccati-Bessel and Riccati-Neumann fucntions $j_\ell(pr)$ and $n_\ell(pr)$ in the presence of the Coulomb potential, respectively. The difference between the two, denoted by $C_\ell(\eta)$, is referred to as the Coulomb barrier factor and is given by
\begin{align}
    C_\ell(\eta) 
    =
    \frac{e^{\pi\eta}}{\Gamma(\ell+1)}
    \exp 
    \left(
    \frac{1}{2}(\ln \Gamma(\ell+1-i\eta)+\ln \Gamma(\ell+1+i\eta))
    \right).
\end{align}
By taking a linear combination of $F_\ell(pr)$ and $G_\ell(pr)$, one can construct the Coulomb–Hankel functions $H_\ell^\pm(pr)$, which correspond to the Riccati–Hankel functions $h_\ell^\pm(pr)$ in the presence of the Coulomb potential:
\begin{align}
    H^\pm_\ell(pr) = G_\ell(pr) \pm i F_\ell(pr).
\end{align}
The asymptotic behavior of $H_\ell^\pm(pr)$ at large distances is the same as that of the Riccati–Hankel function $h_\ell^\pm(pr)$, apart from a logarithmic distortion and a phase shift due to the Coulomb potential,
\begin{align}
    \label{App:Coulomb Hankel at infinity}
    H^\pm_\ell(pr)
    \underset{r \to \infty}{\longrightarrow} e^{\pm i(pr +\eta \ln(2pr)-\ell\pi/2+\sigma_\ell)}.
\end{align}

In order to define the Jost function in the presence of a Coulomb potential , we consider the regular solution $\phi^C_{\ell,p}(r)$ to the Schr\"{o}dinger equation (\ref{App: Coulomb schro}). The regular solution is defined as the one that satisfies the boundary condition $\phi^C_{\ell,p}(r) \to (pr)^{\ell+1}/(2\ell+1)!!$ at the origin $r=0$. Therefore, from Eq.~(\ref{App: Coulomb waves at the origin}), we have
\begin{align}
    \phi^C_{\ell,p}(r) 
    = 
    \frac{1}{C_\ell(\eta)} F_\ell(pr)
    = 
    \frac{i}{2}
    \left[
    \frac{1}{C_\ell(\eta)}H_\ell^-(pr)
    -
    \frac{1}{C_\ell(\eta)} H_\ell^+(pr) 
    \right],
\end{align}
whose asymptotic behavior at infinity is given by
\begin{align}
    \phi^C_{\ell,p}(r)
    \underset{r \to \infty}{\longrightarrow}
    \frac{i}{2}
    \left[
    \frac{e^{-i\sigma_\ell}}{C_\ell(\eta)}e^{-i(pr+\eta \ln(2pr)-\ell \pi/2)}
    -
    \frac{e^{i\sigma_\ell}}{C_\ell(\eta)}e^{i(pr-\eta \ln(2pr)+\ell \pi/2)}
    \right].
\end{align}
From the above expression, the Coulomb Jost functions $\mathscr{J}^{C({\rm in/out})}_\ell(p)$ are defined as the amplitude associated with the plane wave modified by the logarithmic correction at infinity in the regular solution:
\begin{align}
    \mathscr{J}_\ell^{C({\rm in/out})}(p) 
    =
    \frac{e^{\mp i\sigma_\ell}}{C_\ell(\eta)}
    =
    \frac{e^{-\pi \eta}\ell!}{\Gamma(\ell+1\mp i \eta)}.
\end{align}
The $S$-matrix for the $\ell$-th partial wave in Coulomb scattering is given by the ratio of the Jost functions for the incoming and outgoing waves,
\begin{align}
    s_\ell^C(p)
    = \frac{\mathscr{J}_\ell^{C({\rm out})}(p)}{\mathscr{J}_\ell^{C({\rm in})}(p)}
    =
    e^{2i\sigma_\ell}.
\end{align}
Furthermore, since the Gamma function has simple poles at non-positive integers, the Jost function for the incoming wave $\mathscr{J}_\ell^{C({\rm in})}(p)$ exhibits infinitely many zeros located along the positive imaginary axis $p=i\kappa \,(\kappa>0)$ in the case of an attractive Coulomb potential $\alpha >0$, which are determined by
\begin{align}
    \ell +1 -\frac{\mu \alpha}{\kappa} = -n_r, \quad n_r = 0,1,2,\dots.
\end{align}
These correspond to bound states with the following binding energies:
\begin{align}
    E_n = -\frac{\mu \alpha^2}{2n^2}
\end{align}
where $n=n_r+\ell+1$ is the principal quantum number. This is nothing but the spectrum of bound-state energies for the hydrogen atom.

In the presence of a Coulomb potential, the Sommerfeld factor can be expressed, in analogy with Eq.~(\ref{eq:def SE}), as the inverse squared modulus of the Jost function for the incoming wave:
\begin{align}
    S_{\ell}^C(p)
    = 
    \frac{1}{|\mathscr{J}_\ell^{C({\rm in})}(p)|^2}
    =
    \frac{1}{C_\ell(\eta)^2}
    =
    \frac{2\pi \eta}{1-e^{-2\pi \eta}}\prod_{b=1}^{\ell}
    \left(
    1+\frac{\eta^2}{b^2}
    \right)
\end{align}
Here, we have used the identity for the gamma function (valid for real $a$), $|\Gamma(1+ia)|^2=\pi a/ \sinh (\pi a)$.

\section{Several relations involving the Jost function}
\label{app: Jost}

This Appendix discusses several relations concerning the Jost function that were omitted in the main text. In Sec~\ref{app:relation1}, we derive Eq.~(\ref{eq:def jost}) and Eq.~(\ref{eq:low energy expansion of jost}), while Sec~\ref{app:relation2} focuses on Eq.~(\ref{eq:BS 2}). It also presents results from complex analysis based on the fact that the Jost function is an analytic function of momentum, including Levinson's theorem in Sec~\ref{app:Levinson's theorem} and the dispersion relations in Sec~\ref{app:dispersion}. For a more detailed exposition of the material presented here, see\,\cite{Newton:1982qc,Taylor:1972pty}.

\subsection{Relations around Eqs.~(\ref{eq:def jost}) and (\ref{eq:low energy expansion of jost})}
\label{app:relation1}

We begin by presenting an integral representation of the Jost function in Eq.~(\ref{eq:def jost}). The Jost function is defined from the asymptotic behavior of the regular solution to the Schrödinger equation at spatial infinity $r \to \infty$. Therefore our starting point is the Schrödinger equation, taking the mass, the potential, and the momentum to be $\mu$, $V(r)$, and $p > 0$, respectively,
\begin{align}
    \label{App:schro}
    \left[\frac{d^2}{dr^2}-\frac{\ell(\ell+1)}{r^2}-2\mu V(r) +p^2\right]\phi_{\ell,p}(r) = 0.
\end{align}
We construct the regular solution satisfying the boundary condition in Eq.~(\ref{eq:regular bc}) to this equation using the method of variation of parameters. Specifically, we take the Riccati–Hankel functions $h_\ell^\pm (pr)$ as the fundamental solutions in the absence of the potential $V(r)=0$ and express the regular solution as
\begin{align}
    \label{App:variation of parameters}
    \phi_{\ell,p}(r) = \frac{i}{2}\left[J^{\rm in}_\ell(p,r) h_{\ell}^-(pr) - J^{\rm out}_\ell(p,r) h^+_\ell(pr)\right],
\end{align}
with undetermined coefficients $J_\ell^{\rm in/ out}(p,r)$. Since there are two undetermined functions associated with a single differential equation, there exists an inherent ambiguity in their definition. This ambiguity can be fixed by imposing the following condition:
\begin{align}
    \label{App:lagrange condition}
    h^-_\ell(pr)\frac{\partial}{\partial r}J^{\rm in}_\ell(p,r) - h_\ell^+(pr)\frac{\partial}{\partial r}J^{\rm out}_\ell(p,r) = 0.
\end{align}
From the boundary condition imposed on the regular solution, together with Eqs.~(\ref{App:schro}), (\ref{App:variation of parameters}), and (\ref{App:lagrange condition}), the coefficient functions $J_\ell^{\rm in/out}(p,r)$ are found to satisfy the following system of differential equations and boundary conditions:
\begin{align}
    \frac{d}{dr} J^{\rm in}_\ell(p,r) &= \frac{i\mu}{p}h^+_\ell(pr) V(r) [J^{\rm in}_\ell(p,r)h_{\ell}^-(pr)-J_\ell(p,r)^{\rm our}h^+_\ell(pr)] \\
    \frac{d}{dr} J^{\rm out}_\ell(p,r) &= \frac{i\mu}{p}h^-_\ell(pr) V(r) [J^{\rm in}_\ell(p,r)h_{\ell}^-(pr)-J_\ell(p,r)^{\rm our}h^+_\ell(pr)] \\
    & \qquad J^{\rm in}_\ell(p,0) = J^{\rm out}_\ell(p,0) = 1
\end{align}
Alternatively, one obtains the equivalent integral equations:
\begin{align}
    J_\ell^{\rm in/out}(p,r) &= 1+ \frac{2\mu}{p}\int^r_0 dr' h_\ell^\pm(pr')V(r')\phi_{\ell,p}(r') 
\end{align}
Provided that the potential $V(r)$ possesses sufficiently good properties, the integral on the RHS converges as $r \to \infty$, and the Jost functions are obtained as the limiting value of these expressions,
\begin{align}
    \label{App:Jost integral}
    \mathscr{J}^{\rm in/out}_\ell(p) = J^{\rm in/out}_\ell(p,\infty) = 1 + \frac{2\mu}{p}\int^\infty_0 dr h^{\pm}_\ell(pr)V(r)\phi_{\ell,p}(r).
\end{align}
 Furthermore, by deforming the integration contour, the Jost functions originally defined for real positive $p$ can be analytically continued to the complex $p$-plane, although the precise domain of this continuation depends on the detailed properties of the potential. For real values of $p$, the regular solution is real-valued, and under momentum inversion $p \to -p$, both the regular solution and the Riccati-Hankel functions satisfy symmetry relations, $\phi_{\ell,-p}(r) = (-1)^{\ell+1}\phi_{\ell,p}(r)$ and $h^+_\ell(-pr) = (-1)^{\ell}h_\ell^-(pr)^*$. These two facts together imply for real $p$,
\begin{align}
    \label{App:Jost conjugation}
    \mathscr{J}^{\rm in}_\ell(-p) = \mathscr{J}_\ell^{\rm out}(p) = [\mathscr{J}_\ell^{\rm in}(p)]^*
\end{align}
From the initial equality, the two analytic functions $\mathscr{J}^{\rm in}_\ell(-p)$ and $\mathscr{J}^{\rm out}_\ell(p)$ coincide on the real axis in $p$, and hence, by analytic continuation, they must be equal throughout the complex $p$-plane. If we denote $\mathscr{J}^{\rm in}_\ell(p)$ simply by $\mathscr{J}_\ell(p)$, it follows that $\mathscr{J}_\ell^{\rm out}(p) = \mathscr{J}_\ell(-p)$ and the integral representation for $\mathscr{J}_\ell^{\rm in}(p)$ demonstrates Eq.~(\ref{eq:def jost}).

From the integral representation~(\ref{App:Jost integral}), one can investigate the structure of the $p$-dependence of the Jost function. By decomposing the Riccati--Hankel function $h_\ell^+(pr)$ into the Riccati--Bessel and Riccati--Neumann functions, $j_\ell(pr)$ and $n_\ell(pr)$, Eq.~(\ref{App:Jost integral}) can be rewritten as
\begin{align}
    \mathscr{J}_\ell(p) = 1 + \frac{1}{p}\int^\infty_0 dr\, n_\ell(pr) V(r) \phi_{\ell,p}(r) + i \frac{1}{p}\int^\infty_0 dr \,j_\ell(pr) V(r) \phi_{\ell,p}(r)
\end{align}
It follows from Appendix~\ref{app: Bessel} that $n_\ell(pr)$ exhibits $p^{-\ell}$ times a power series in $p^2$, while $j_\ell(pr)$ exhibits $p^{\ell+1}$ times a power series in $p^2$. On the other hand, regarding the $p$-dependence of the regular solution $\phi_{\ell,p}(r)$, the Schr\"{o}dinger equation Eq.~(\ref{App:schro}) is analytic in $p^2$, and therefore the solution has the form of a $p^{\ell+1}$ factor times a power series in $p^2$, as dictated by the boundary condition. Thus we conclude that the Jost function has the following form
\begin{align}
    \label{App:Jost low energy}
    \mathscr{J}_\ell(p) = 1 + [\alpha_\ell + \beta_\ell p^2 + \mathcal{O}(p^4)] + ip^{2\ell+1}[\gamma_\ell + \mathcal{O}(p^2)],
\end{align}
with all coefficients being real, and can be expressed as in Eq.~(\ref{eq:low energy expansion of jost}).

\subsection{Proof of the differentiation formula for the Jost function}
\label{app:relation2}

We derive the formula for the derivative of the Jost function at the bound-state zero, as given in Eq.~(\ref{eq:BS 2}).
To this end, we introduce the so-called Jost solutions $\chi^\pm_{\ell,p}(r)$, which are defined as the solutions to the Schr\"{o}dinger equation satisfying the following boundary conditions:
\begin{align}
    \chi^{\pm}_{\ell,p}(r)/h^{\pm}_\ell(pr) \to 1 \quad r \to \infty.
\end{align}
That is, the Jost solutions are the solutions that asymptotically coincide with the Riccati–Hankel functions at spatial infinity. Since they form a pair of linearly independent solutions to Eq.~(\ref{App:schro}), any solution can be written as a linear combination of them. The Jost function is then defined as the coefficient in the expansion of the regular solution in terms of the Jost solutions:
\begin{align}
    \phi_{\ell,p}(r) = \frac{i}{2}[\mathscr{J}_\ell(p)\chi^-_{\ell,p}(r)-\mathscr{J}_\ell(-p)\chi^+_{\ell,p}(r)]
\end{align}
Alternatively, using the Wronskian, the Jost function can be written explicitly as
\begin{align}
    \label{App:Jost Wronskian}
    \mathscr{J}_\ell(p) = \frac{1}{p}W[\chi^+_{\ell,p},\phi_{\ell,p}].
\end{align}
Here the Wronskian $W[f,g]$ of two functions $f(r)$ and $g(r)$ is defined by
\begin{align}
    W[f,g] = f(r)g'(r) - f'(r)g(r).
\end{align}
By differentiating the expression (\ref{App:Jost Wronskian}) with respect to the momentum $p$, one can directly evaluate the derivative of the Jost function. At the momentum corresponding to a bound state $p=p_B$, $\mathscr{J}_\ell(p_B)$ vanishes and the regular solution becomes proportional to the Jost solution as $\phi_{\ell,p_B} = -i\mathscr{J}_\ell(p_B)\chi^+_{\ell,p_B}/2$, from which we obtain
\begin{align}
    \left.
    \frac{d}{dp}\mathscr{J}_\ell(p)
    \right|_{p=p_B} 
    =
    \frac{1}{p_B}
    \left.
    \left
    (W\left[\frac{\partial \chi^+_{\ell,p}}{\partial p},\phi_{\ell,p}\right]
    +
    W\left[\chi^+_{\ell,p},\frac{\partial \phi_{\ell,p}}{\partial p}\right]
    \right)
    \right|_{p=p_B}.
\end{align}
To evaluate the Wronskian on the RHS, we observe that the $r$-derivative of the Wronskian $W[\chi^+_{\ell,p_B},\phi_{\ell,p}]$ is given, via the Schr\"{o}dinger equation in Eq.~(\ref{App:schro}), by
\begin{align}
    \frac{\partial}{\partial r}W[\chi^+_{\ell,p_B},\phi_{\ell,p}] = (p_B^2-p^2)\chi^+_{\ell,p_B}\phi_{\ell,p},
\end{align}
which, upon further differentiation with respect to $p_B$ and $p$, yields
\begin{align}
    \frac{\partial}{\partial r} W
    \left[
    \frac{\partial{\chi^+_{\ell,p_B}}}{\partial p_B},\phi_{\ell,p_B} 
    \right]
    =
    -2p_B \chi^+_{\ell,p_B}\phi_{\ell,p_B},
    \quad
    \frac{\partial}{\partial r} W
    \left[
    \chi^+_{\ell,p_B},\frac{\partial \phi_{\ell,p_B}}{\partial p_B}
    \right]
    =
    2p_B \chi^+_{\ell,p_B}\phi_{\ell,p_B}
\end{align}
By integrating the first equation from 0 to $r$, and the second from $r$ to $\infty$, and using the fact that both the regular solution $\phi_{\ell,p_B}(r)$ and the Jost solution $\chi^+_{\ell,p_B}(r)$ at a bound state momentum $p=p_B$ are regular at the origin and vanish at infinity, the corresponding integration constants vanish, leading to
\begin{align}
    \left.
    \left
    (W\left[\frac{\partial \chi^+_{\ell,p}}{\partial p},\phi_{\ell,p}\right]
    +
    W\left[\chi^+_{\ell,p},\frac{\partial \phi_{\ell,p}}{\partial p} \right]
    \right)
    \right|_{p=p_B}
    =i \mathscr{J}_\ell(-p_B) \int^\infty_0 dr (\chi_{\ell,p_B}(r))^2.
\end{align}
By comparing the asymptotic behavior of the Jost solution at infinity, $\chi^{+}_{\ell,p_B}(r)\to (-i)^\ell e^{ip_B r}$, with that of the normalized and real-valued reduced wave function for the bound state, $u_{\ell,p_B}(r) \to \mathcal{A}_\ell e^{ip_B r}$, one finds that the two are related by
\begin{align}
    u_{\ell,p_B}(r) &= i^\ell \mathcal{A}_\ell \chi_{\ell,p_B}(r).
\end{align}
This, when combined with the result above, leads to Eq.~(\ref{eq:BS 2}). Note that the above derivation shows that the derivative of the Jost function at the bound-state momentum is nonzero. Therefore, the bound state corresponds to a simple zero of the Jost function, or equivalently, to a simple pole of the scattering amplitude. However, this is not applicable to the zero-energy resonance, because in that case the wave function does not decay at infinity and the integral does not converge. In fact, from the low-energy expansion of the Jost function (\ref{App:Jost low energy}), the zero-energy resonance corresponds to a simple zero for the $s$-wave, whereas for the $p$-wave and higher partial waves, it becomes a double zero.

\subsection{Levinson's theorem}
\label{app:Levinson's theorem}

As an application that combines the Jost function with complex analysis, we present a proof of Levinson’s theorem. Levinson’s theorem states that the phase shift $\delta_\ell(p)$ is related to the number of bound states in the system:
\begin{align}
    \label{App: Levinson theorem}
    \delta_\ell(0)-\delta_\ell(\infty) = \left(n_\ell + \frac{N}{2}\right) \pi,
\end{align}
where $n_\ell$ denotes the number of non-zero-energy bound states in the $\ell$-th partial wave, and the extra factor $N$ accounts for the presence of a zero-energy resonance. Specifically, $N = 0$ in the absence of a zero-energy resonance, and takes the value $1$ or $2$ for the $s$-wave and higher partial waves, respectively, when such a resonance is present. In particular, taking into account the assumption $\delta_\ell(\infty) = 0$ used in the main text, Levinson’s theorem implies that the value of the phase shift at zero momentum is closely related to the number of bound states.

To prove this theorem, we consider the following complex contour integral:
\begin{align}
    \label{App:Levinson}
    I = \oint_C\, \frac{dp}{\mathscr{J}_\ell(p)}\frac{d}{dp}\mathscr{J}_\ell(p)
\end{align}
where the integration contour $C$ is taken as a semicircle in the upper half of the complex $p$-plane that encloses the real axis. However, when a zero-energy resonance is present in the spectrum, the integrand develops a singularity at $p = 0$, and thus the contour must detour around the origin along a small semicircle of radius $\varepsilon$, which is denoted by $C_\varepsilon$. The only singularities of the integrand inside the contour $C$ are simple poles originating from the bound states, each of which contributes a residue of 1. Therefore, by the residue theorem, $I$ is given by
\begin{align}
    \label{App:residue}
    I = 2\pi i n_\ell.
\end{align}

On the other hand, the contribution from the arc of the contour integral vanishes in the limit where the radius of the semicircle is taken to infinity. Therefore, the only contributions to $I$ come from the integrals along the real axis and the small semicircle $C_\varepsilon$:
\begin{align}
    I = \lim_{\varepsilon \to 0} 
    \left(
    \int_{-\infty}^{-\varepsilon} + \int_\varepsilon^\infty+
    \int_{C_\varepsilon} 
    \right )
    \frac{dp}{\mathscr{J}_\ell(p)}\frac{d}{dp}\mathscr{J}_\ell(p).
\end{align}
As for the integral along the real axis, the integrand can be written as $d \ln \mathscr{J}_\ell(p)$. For $p > 0$, it satisfies $\ln \mathscr{J}_\ell(p) = \ln |\mathscr{J}_\ell(p)|-i\delta_\ell(p)$, while for $p < 0$, it follows from Eq.~(\ref{App:Jost conjugation}) that $\ln \mathscr{J}_\ell(-p)=\ln|\mathscr{J}_\ell(p)| + i \delta_\ell(p)$. Therefore,
\begin{align}
    \label{App:real axis}
    \lim_{\varepsilon \to 0} 
    \left(
    \int_{-\infty}^{-\varepsilon} + \int_\varepsilon^\infty
    \right )
    \frac{dp}{\mathscr{J}_\ell(p)}\frac{d}{dp}\mathscr{J}_\ell(p)
    = -2i\int^\infty_0 d \delta_\ell(p) = 2i[\delta_\ell(0)-\delta_\ell(\infty)] .
\end{align}
The integral over the small semicircle $C_\varepsilon$ contributes only when a zero-energy resonance is present. In such a case, the Jost function has a simple zero for the $s$-wave and a double zero for higher partial waves, yielding
\begin{align}
    \label{App:small semi circle}
    \lim_{\varepsilon \to 0} 
    \int_{C_\varepsilon} 
    \frac{dp}{\mathscr{J}_\ell(p)}\frac{d}{dp}\mathscr{J}_\ell(p)
    =
    -N \pi i .
\end{align}
By combining Eqs.~(\ref{App:residue}), (\ref{App:real axis}), and (\ref{App:small semi circle}), Levinson’s theorem (\ref{App: Levinson theorem}) follows.

\subsection{Dispersion relation}
\label{app:dispersion}

In this subsection, we consider the dispersion relation for the Jost function. As a starting point, let us examine the simplest case of a dispersion relation for a complex function $f(z)$ that is analytic in the upper half-plane. According to Cauchy’s integral theorem, if the contour $C$ encloses the upper half-plane including the real axis, the following identity holds:
\begin{align}
    f(z) = \frac{1}{2\pi i} \oint_C dz'\, \frac{f(z')}{z'-z}, \quad {\rm Im}\,z > 0.
\end{align}
If $f(z)$ vanishes sufficiently fast as $|z| \to \infty$, then the contribution from the arc of the contour integral becomes zero in the limit of an infinitely large radius of $C$. Therefore, for $z = x+i0^+$ with real $x$, we obtain
\begin{align}
    \label{App:DR 1}
    f(x) = \frac{1}{2\pi i} \int^\infty_{-\infty} dx'\,\frac{f(x')}{x'-x-i0^+} = \frac{1}{\pi i} \mathcal{P} \int^\infty_{-\infty} dx' \frac{f(x')}{x'-x}.
\end{align}
To derive the second equality, we have used the Sokhotski–Plemelj formula $(x-i0^+)^{-1}=\mathcal{P}(1/x)+i\pi \delta(x)$. Here, $\mathcal{P}$ denotes the Cauchy principal value. By examining the real and imaginary parts of both sides of Eq.~(\ref{App:DR 1}), we arrive at the following dispersion relations:
\begin{align}
    \label{App:DR 2}
    {\rm Re}\,f(x) 
    =
    \frac{1}{\pi} \mathcal{P} \int^\infty_{-\infty} dx' \frac{{\rm Im}\,f(x')}{x'-x},
    \quad
    {\rm Im}\,f(x)
    =
    -\frac{1}{\pi} \mathcal{P} \int^\infty_{-\infty} dx' \frac{{\rm Re}\,f(x')}{x'-x}.
\end{align}
The integral operation appearing on the right-hand side of the above equation is known as the Hilbert transform. Accordingly, the dispersion relation states that the real and imaginary parts of an analytic function are connected to each other via the Hilbert transform. Another useful form of the dispersion relation can also be derived from Eq.~(\ref{App:DR 2}) as
\begin{align}
    \label{App:DR 3}
    f(x) = \frac{1}{\pi} \int^\infty_{-\infty}dx'\, \frac{{\rm Im}\,f(x')}{x'-x-i0^+}.
\end{align}

Since the Jost function $\mathscr{J}_\ell(p)$ is an analytic function of the momentum $p$, the above discussion on dispersion relations can be applied: naively speaking, the real part of the logarithm of the Jost function, $\ln \mathscr{J}_\ell(p) = \ln |\mathscr{J}_\ell(p)|-i\delta_\ell(p)$, is given by the Hilbert transform of its imaginary part, which corresponds to the phase shift $\delta_\ell(p)$. Since the modulus of the Jost function gives the Sommerfeld factor and its phase shift carries information about self-scattering, the dispersion relation consequently encodes the correlation between them\,\cite{Kamada:2023iol}. 

To make this argument more precise, let us consider the following auxiliary function:
\begin{align}
\mathscr{I}_\ell(p) = \prod_{n=1}^{n_\ell}\left(\frac{p+i\kappa_n}{p-i\kappa_n}\right)\,\mathscr{J}_\ell(p),
\end{align}
Here, the extra factors are introduced to cancel the zeros of the Jost function at the bound-states $p=p_B =i\kappa_n$ (with $\kappa_n>0$) without altering the asymptotic behavior at infinity. The indices $n=1,\dots,n_\ell$ label the bound states for the $\ell$-th partial wave. Assuming that the system has no zero-energy resonance, the logarithm of $\mathscr{I}_\ell(p)$ is analytic in the upper half of the complex $p$-plane and vanishes as $|p|\to \infty$ under the normalization $\delta_\ell(\infty)=0$, and for real $p$ it becomes
\begin{align}
    \label{App:DR 4}
    \ln \mathscr{I}_\ell(p) 
    =
    \ln |\mathscr{J}_\ell(p)| -i\delta_\ell(p) + i\sum_{n=1}^{n_\ell}\, {\rm Im}\, [\ln(p+i\kappa_n)-\ln (p-i\kappa_n)]
\end{align}
It follows from the dispersion relation in Eq.~(\ref{App:DR 3}) that the following expression holds:
\begin{align}
    \label{App:DR 5}
    \ln \mathscr{I}_\ell(p) 
    =
    \frac{1}{\pi} \int^\infty_{-\infty} dp'\,\frac{1}{p'-p-i0^+}
    \left(
    -\delta_\ell(p') + \sum_{n=1}^{n_\ell}\, {\rm Im}\, [\ln(p'+i\kappa_n)-\ln (p'-i\kappa_n)]
    \right)
\end{align}
The second integral in the above expression can be evaluated using the formula
\begin{align}
    \ln\left |\frac{p\pm i\kappa_n}{p}\right |
    =
    \pm \frac{1}{\pi} \mathcal{P}\int^\infty_{-\infty} dp' \, \frac{{\rm Im}\,[\ln (p' \pm i \kappa_n)-\ln p']}{p'-p}
\end{align}
which is derived from a dispersion relation of the type (\ref{App:DR 2}) for analytic functions $ \ln (p\pm i\kappa_n/p)$ that are regular in the upper (or lower) half-plane. Therefore, by combining Eqs.~(\ref{App:DR 4}) and (\ref{App:DR 5}), we obtain the following relation:
\begin{align}
    \label{App:DR 6}
    \mathscr{J}_\ell(p) 
    =
    \prod_{n=1}^{n_\ell}
    \left(1-\frac{E_n}{E}\right)
    \exp
    \left[
    -\frac{1}{\pi}\int^\infty_{-\infty}dp'\,\frac{\delta_\ell(p')}{p'-p-i0^+}
    \right],
\end{align}
where $E=p^2/2\mu$ and $E_n = -\kappa_n^2/2\mu$ are the energies of the scattering state and $n$-th bound state, respectively. Thus, Eq.~(\ref{App:DR 6}) provides a representation of the Jost function in terms of the bound state energies and the scattering phase shift, valid for real values of $p$.

In deriving the dispersion relation (\ref{App:DR 6}) for the Jost function, we assumed the absence of a zero-energy resonance. Nevertheless, the expression remains valid and reproduces the correct behavior even when such a resonance is present. In particular, in the limit $p\to0$, it yields
\begin{align}
    \mathscr{J}_\ell(p) \sim p^{2(\delta_\ell(0)/\pi-n_\ell)} = p^N,
\end{align}
where Levinson’s theorem (\ref{app:Levinson's theorem}) has been used. As in section~(\ref{App:Levinson}), the exponent $N$ takes values $0$, $1$, or $2$, corresponding to the absence of a zero-energy resonance, its presence in the $s$-wave, and in higher partial waves, respectively. Therefore, the Jost function obtained from Eq.~(\ref{App:DR 6}) exhibits the expected behavior even in the presence of a zero-energy resonance.

\section{Numerical comparison with the direct Schr\"{o}dinger approach}
\label{app: numerical}

In this appendix, we numerically compare the RG method discussed in the main text with the approach commonly found in the literature \cite{Blum:2016nrz, Parikh:2024mwa}, where the annihilation effect is modeled by the delta function and its derivatives, and the Schr\"{o}dinger equation is solved directly to compute the Sommerfeld enhancement in a way consistent with unitarity.
The comparison is performed for the spherical well potential in Eq.~(\ref{eq:well}) for the $\ell=0$ ($s$-wave) and $\ell=1$ ($p$-wave) cases.

First, we briefly outline the formulation of the direct solution method of the Schr\"{o}dinger equation; see \cite{Parikh:2024mwa} for details.
As discussed in the main text, the Sommerfeld enhancement arises in systems where both a long-range potential $V_L(r)$ and a short-range potential $V_S(r)$ are present.
However, since the contact interaction $V_S(r)$ is represented by a delta-function-like potential and thus highly singular, one must regularize it in order to solve the Schr\"{o}dinger equation non-perturbatively.
To this end, we introduce a cutoff scale $r = a$ in position space and define the total potential as follows:
\begin{align}
\label{app: potential reg}
V(r) =
\left \{
\begin{aligned}
&V^{\rm reg}_S(r) \quad (r<a) \\
&V_L(r) \quad (r>a)
\end{aligned}
\right.,
\end{align}
where, in the region shorter than the cutoff, the potential is regularized and represented by the non-singular form $V_S^{\rm reg}(r)$.
Let us now construct the solution of the Schrödinger equation under the potential given in Eq.~(\ref{app: potential reg}).
In $r > a$, the potential reduces to $V_L(r)$ only. We therefore choose two independent solutions in this long-range region—the regular solution $F_{\ell,p}(r)$ and the irregular solution $G_{\ell,p}(r)$—which satisfy the following boundary conditions at the origin\footnote{In \cite{Parikh:2024mwa}, the notation $C_\ell(p) = |\mathscr{J}_\ell(p)|^{-1}$ is adopted.}:
\begin{align}
F_{\ell,p}(r)
\underset{r \to 0}{\longrightarrow}
\frac{1}{|\mathscr{J}_\ell(p)|}\frac{(pr)^{\ell+1}}{(2\ell+1)!!},
\quad
G_{\ell,p}(r)
\underset{r \to 0}{\longrightarrow}
|\mathscr{J}_\ell(p)|\frac{(2\ell-1)!!}{(pr)^\ell},
\end{align}
with $\mathscr{J}_\ell(p)$ being the Jost function of $V_L(r)$. However, since adding a constant multiple of the regular solution to the irregular solution still satisfies the same boundary condition at the origin, a condition at infinity is also necessary to completely fix the solution. Here we impose:
\begin{align}
G_{\ell,p}(r) + iF_{\ell,p}(r)
\underset{r \to \infty}{\longrightarrow}
(-i)^\ell \exp(ipr + \delta^L_\ell(p)),
\end{align}
with $\delta_\ell^L(p)$ being the phase shift due to $V_L(r)$, as in the main text. 

Since the pair of linearly independent solutions has been determined, the solution to the Schrödinger equation in the region outside the cutoff can be expressed as their linear combination. The full wavefunction is then given by
\begin{align}
\label{app: total wave function}
u_{\ell,p}(r)=
\left\{
    \begin{aligned}
        & u_{<,\ell,p}(r) \quad (r<a)\\
        & e^{i(\delta_\ell^L(p) + \delta_\ell^S(p))}(F_{\ell,p}(r)\cos \delta_\ell^S(p) + G_{\ell,p}(r)\sin \delta^S_\ell(p)) \quad (r>a)
    \end{aligned}
\right.,
\end{align}
where $u_{<,\ell,p}(r)$ denotes the wavefunction in the region inside the cutoff, which is regular at the origin and whose explicit form depends on the details of the regularization of the short-range potential $V^{\rm reg}_S(r)$. The complex parameter $\delta_\ell^S(p)$ corresponds to the phase shift induced by the short-range interaction, as seen from the $S$-matrix $s_\ell(p)$ obtained from the asymptotic behavior of the wavefunction in Eq.~(\ref{app: total wave function}), which takes the form $s_\ell(p) = \exp(2i(\delta^L_\ell(p) + \delta^S_\ell(p)))$. To facilitate the discussion, it is convenient to introduce the quantity $k_\ell(p)$ defined as follows:
\begin{align}
k_\ell(p) 
=
-\frac{p^{2{\ell+1}}}{|\mathscr{J}_\ell(p)|^2\tan \delta^S_\ell(p)}
=
\frac{p^{2\ell+1}}{|\mathscr{J}_\ell(p)|^2}
\left.
  \left(
    \frac{G'_{\ell,p}(r)-G_{\ell,p}(r)\,u'_{<,\ell,p}(r)/u_{<,\ell,p}(r)}
    {F'_{\ell,p}(r)-F_{\ell,p}(r)\,u'_{<,\ell,p}(r)/u_{<,\ell,p}(r)}
  \right)
\right|_{r=a}.
\end{align}
The equality on the far right-hand side indicates that $\delta_\ell^S(p)$ is determined by the matching condition of the wavefunction at the cutoff. Using $k_\ell(p)$, the $S$-matrix $s_\ell(p)$ or the self-scattering amplitude $f_\ell(p)=(s_\ell(p)-1)/2ip$ can be written as follows:
\begin{align}
s_\ell(p) 
&
= \frac{k_\ell(p)-ip^{2\ell+1}/|\mathscr{J}_\ell(p)|^2}{k_\ell(p)+ip^{2\ell+1}/|\mathscr{J}_\ell(p)|^2}
e^{2i\delta^L_\ell(p)}, \\
f_\ell(p)
&=
f^L_\ell(p) - \frac{1}{\mathscr{J}_\ell(p)^2}\frac{p^{2\ell}}{k_\ell(p)+ip^{2\ell+1}/|\mathscr{J}_\ell(p)|^2},
\end{align}
where $f_\ell^L(p)$ is the scattering amplitude due to the long-range potential $V_L(r)$. Therefore, the scattering problem for the total potential is reduced to the determination of $k_\ell(p)$. By using the behavior of solutions $F_{\ell,p}(r)$ and $G_{\ell,p}(r)$ near the origin and assuming a momentum dependence of the regularization-dependent quantity $u'_{<,\ell,p}(a)/u_{<,\ell,p}(a)$, Ref.~\cite{Parikh:2024mwa} gives the approximate momentum dependence of $k_\ell(p)$ as follows:
\begin{align}
\label{app: k function}
k_\ell(p) &\simeq k_\ell(p_0) + \Delta z_\ell(p,p_0;a), \\
\label{app: delta z}
\Delta z_\ell(p,p_0;a) &= \tilde{z}_\ell(p;a) - \tilde{z}_\ell(p_0;a),\\
\label{app: tilde z}
\tilde{z}_\ell(p;a) &=
\left.
  \left[
    \frac{p^\ell}{2^\ell \ell!|\mathscr{J}_\ell(p)|}\frac{d^{2\ell+1}}{dr^{2\ell+1}}(r^\ell G_{\ell,p}(r))
  \right]
\right|_{r=a},
\end{align}
with $p_0$ being some reference momentum. Then the optical theorem in Eq.~(\ref{eq:optical annhilation}) reads the Sommerfeld factor
\begin{align}
\label{app: SE via schro}
S_\ell(p) 
=
\left|
  \mathscr{J}_\ell(p)+\frac{1}{k_\ell(p_0)}
  \left(
    \frac{\Delta z_\ell(p,p_0;a)}{\mathscr{J}_\ell(p)}+ip^{2\ell+1}
  \right)
\right|^{-2}.
\end{align}
Focusing on the regime where the contribution of the short-range potential to the scattering can be treated perturbatively, we can determine $k_\ell(p_0)$ at the matching scale $p_0$. In the original reference \cite{Parikh:2024mwa}, this operation is performed at the level of the cross section; here, by employing the leading-order DWBA in Eq.~(\ref{eq:DWBA simplified}) and the perturbative UV amplitude $f_\ell^S(p)$, we carry it out at the amplitude level:
\begin{align}
\label{appk matching}
\frac{1}{k_\ell(p_0)} 
=
\left(
  -\frac{p_0^{2\ell}}{f_\ell^S(p_0)} -ip_0^{2\ell+1}
\right)^{-1}
\simeq
-i\frac{\mu (\sigma_\ell v)^{\rm ann,0}} { 4\pi(2\ell+1)p_0^{2\ell}},
\end{align}
In the last equality we approximate the UV amplitude by its absorptive part, namely the annihilation cross section.
\begin{figure}[t]
    \centering
    \includegraphics[width=0.49\linewidth]{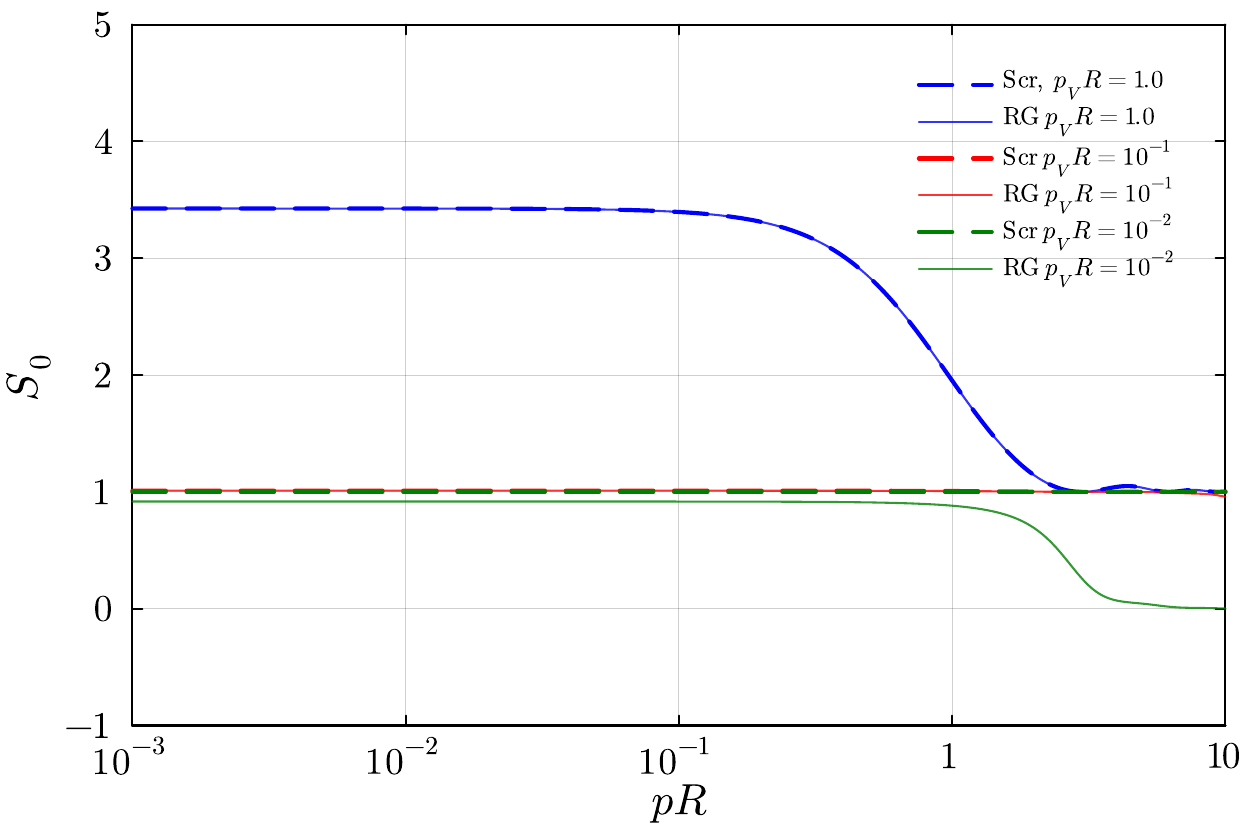}
    \includegraphics[width=0.49\linewidth]{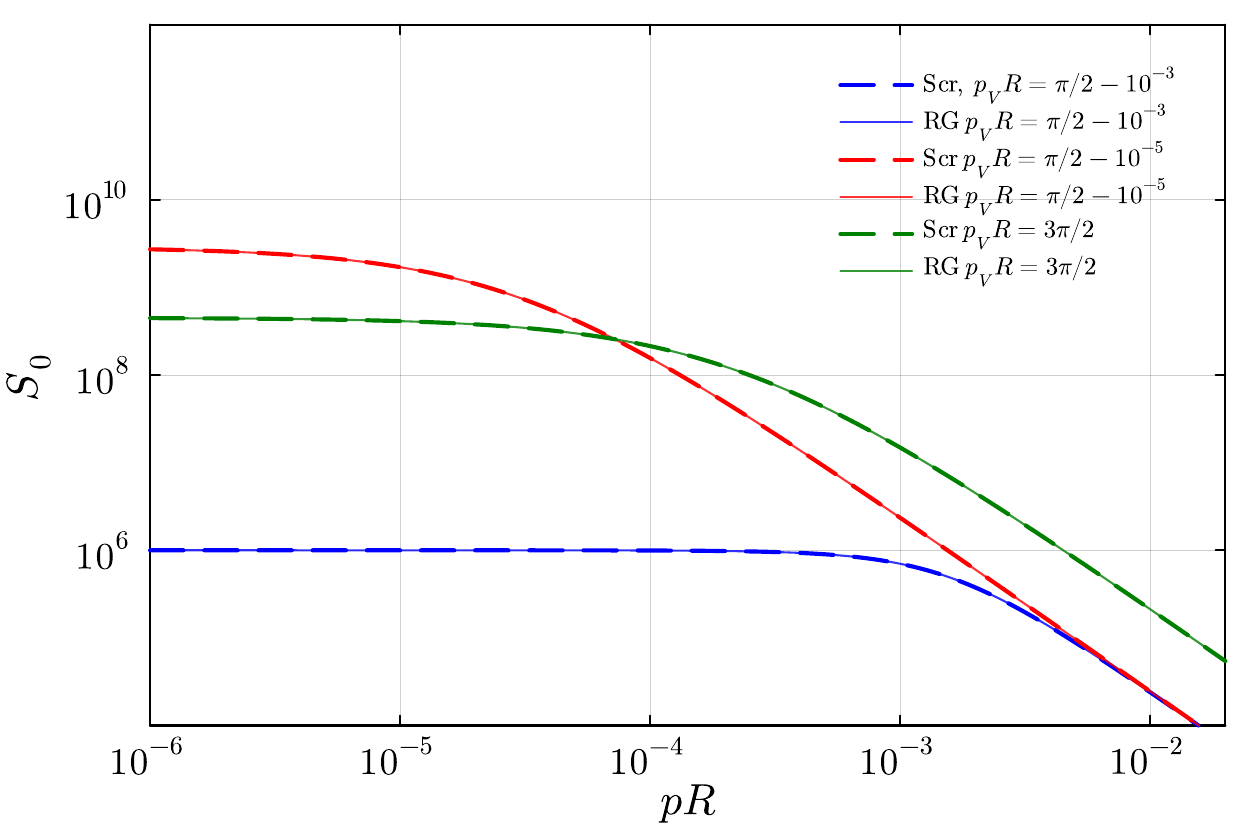}
    \caption{\small \sl 
   Comparison of the unitarized s-wave Sommerfeld factor computed with the RG method (RG; solid) and by solving the Schr\"{o}dinger equation (Scr; dashed). The left panel uses well depths $p_VR = 1,\,10^{-1},\,10^{-2}$; the right panel uses $p_VR = \pi/2-10^{-3},\,\pi/2-10^{-5},\,3\pi/2$. The mass parameter and coupling are fixed to $\mu R = 10$ and $\alpha_D = 10^{-2}$, respectively.
    }
    \label{fig: comparison s-wave}
\end{figure}

\begin{figure}[t]
    \centering
    \includegraphics[width=0.495\linewidth]{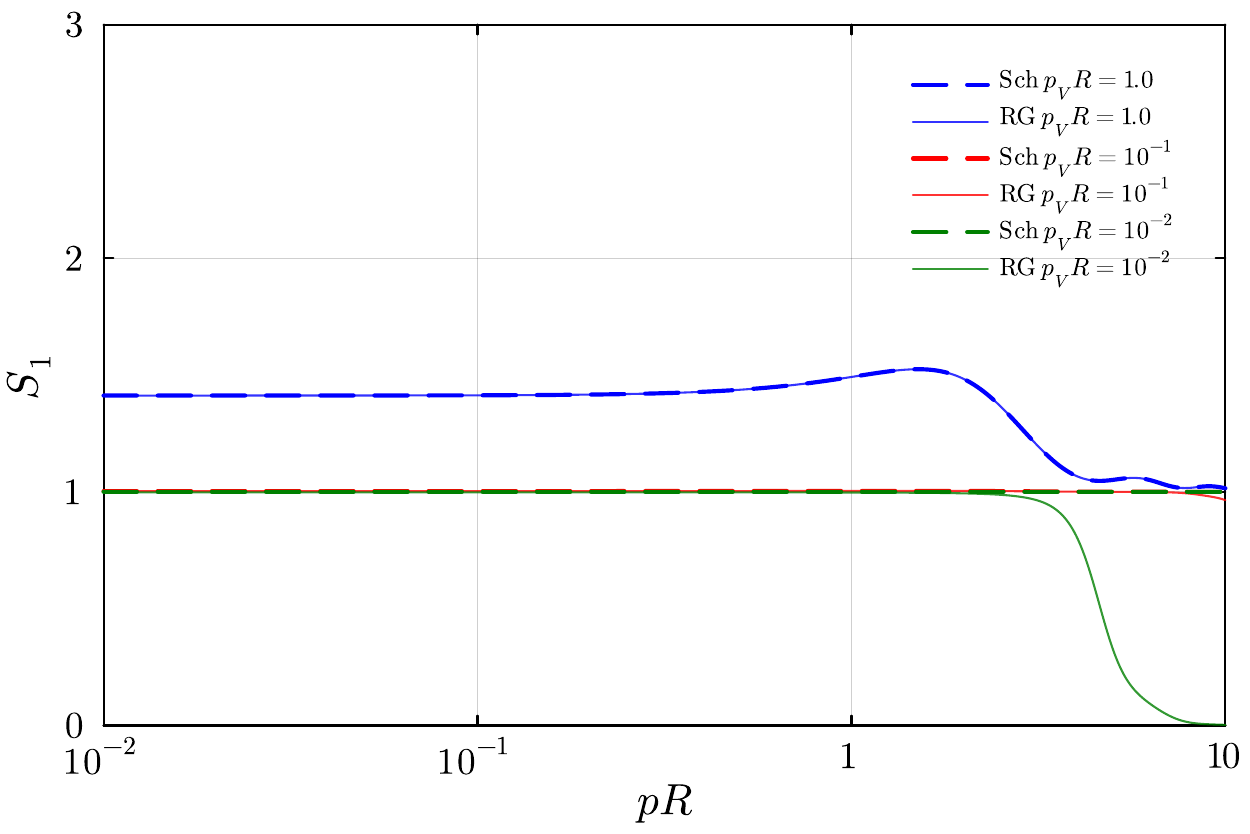}
    \includegraphics[width=0.495\linewidth]{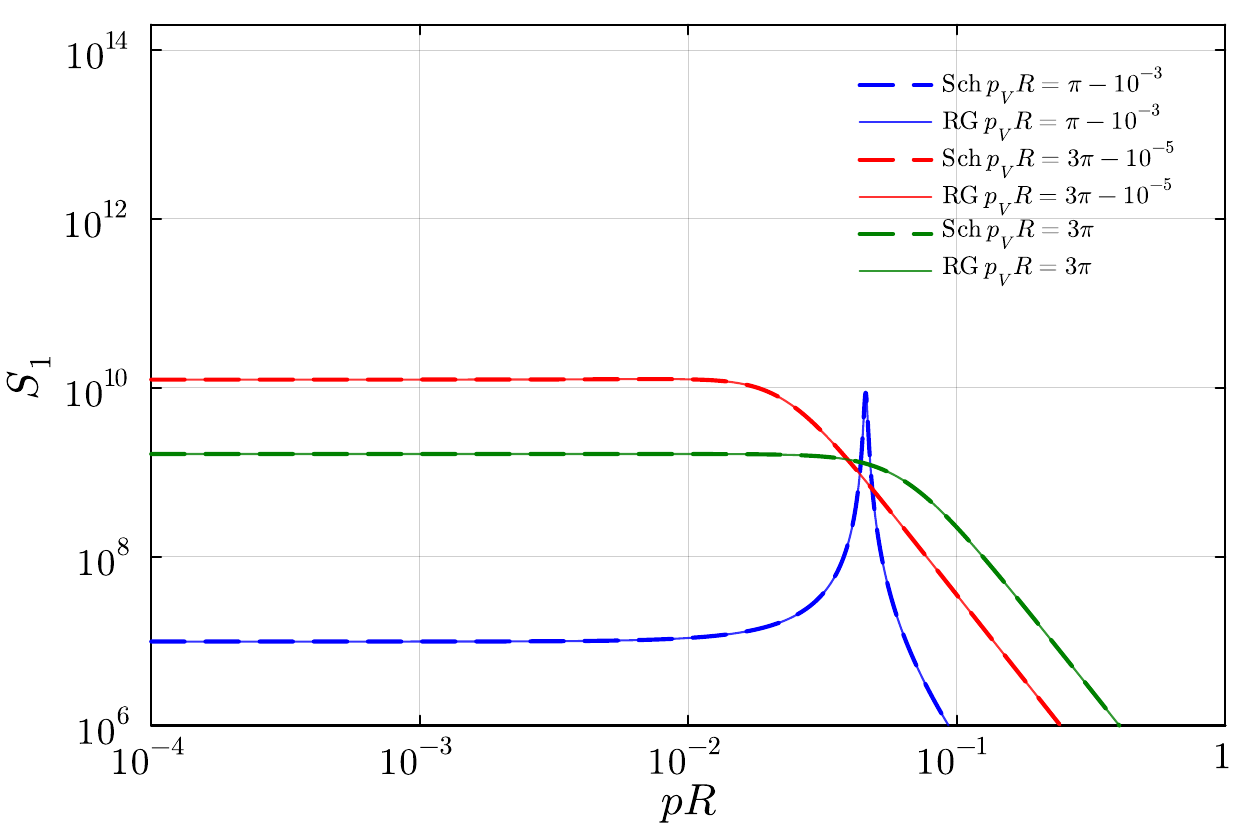}
    \caption{\small \sl 
    Comparison of the unitarized p-wave Sommerfeld factor computed with the RG method (RG; solid) and by solving the Schr\"{o}dinger equation (Scr; dashed). The left panel uses well depths $p_VR = 1,\,10^{-1},\,10^{-2}$; the right panel uses $p_VR = \pi-10^{-3},\,\pi-10^{-5},\,3\pi$. The mass parameter and coupling are fixed to $\mu R = 10$ and $\alpha_D = 10^{-2}$, respectively.
    }
    \label{fig: comparison p-wave}
\end{figure}

Based on the above setup, we numerically compare the improved Sommerfeld factor, consistent with unitarity, computed via the RG method with that obtained by directly solving the Schr\"{o}dinger equation, assuming the long-range potential is the spherical well in Eq.~(\ref{eq:well}) and the UV amplitude is given by Eq.~(\ref{eq:tree level annihilation}). To apply the formula Eq.~(\ref{app: SE via schro}), the $\tilde{z}_\ell$ function in Eq.~(\ref{app: tilde z}) is required, which can be obtained by 
\begin{align}
\lim_{a \to 0}\tilde{z}_\ell(p;a) = -p^\ell \tilde{p}^{\ell+1}\frac{{\rm Im}\,c_F}{|\mathscr{J}_\ell(p)|}.
\end{align}
Here, for the case of the spherical well potential, we used the fact that the cutoff $a$ can be removed without any obstruction. The function $c_F$ is given by the following\footnote{The Riccati–Bessel function $j_\ell(x)$ and the Riccati–Neumann function $n_\ell(x)$ are denoted respectively by $s_\ell(x)$ and $c_\ell(x)$ in \cite{Parikh:2024mwa}.}:
\begin{align}
c_F = \frac{e^{i\delta_\ell^L}}{\tilde{p}}
\Big[
  i\tilde{p}n_\ell'(\tilde{p}R)(n_\ell(pR)+ij_\ell(pR))
  -ip n_\ell(\tilde{p}R)(n'_\ell(pR)+ij_\ell'(pR))
\Big].
\end{align}
Fig.~\ref{fig: comparison s-wave} compares the Sommerfeld factor for the $s$-wave, $S_0(p)$, computed consistently with unitarity via the RG method (solid lines) and by solving the Sch\"{o}dinger equation (dashed lines) under the spherical well potential. The left panel illustrates cases where the conventional Sommerfeld factor is applicable—well depths away from the zero-energy resonance—using $p_VR = 1.0$ (blue), $10^{-1}$ (red), and $10^{-2}$ (green). The right panel focuses on parameters near the appearance of the zero-energy resonance, with $p_VR = \pi/2-10^{-3}$ (blue), $\pi/2-10^{-5}$ (red), and $3\pi/2$ (green). We used the parametrization for the UV amplitude in Eq.~(\ref{eq:tree level annihilation}) and set the mass parameter, the coupling constant and the reference momentum $p_0$ in Eq.~(\ref{app:})to $\mu R = 10$, $\alpha_D = 10^{-2}$ and $p_0R = 10$, respectively. We also present in Fig.~\ref{fig: comparison p-wave} the comparison for the $p$-wave case. The left panel shows off-resonance well depths $p_VR = 1.0$ (blue), $10^{-1}$ (red), and $10^{-2}$ (green), while the right panel shows well depths near the zero-energy resonance $p_VR = \pi-10^{-3}$ (blue), $3\pi-10^{-5}$ (red), and $3\pi$ (green). The two methods agree extremely well across the parameter region—including near the zero-energy resonance—where the Sommerfeld effect becomes sizable. Discrepancies arise only where the long-range interaction is very small and the Sommerfeld contribution is essentially negligible. As discussed in the main text, this occurs because the short-range perturbation to the amplitude can no longer be renormalized into the long-range part. In such cases, however, the Sommerfeld effect is absent, so this mismatch is immaterial from a practical perspective. We therefore conclude that the RG method works adequately in precisely those situations where a Sommerfeld enhancement is present.
\bibliographystyle{unsrt}
\bibliography{references}

\end{document}